\RequirePackage{fix-cm}
\documentclass[smallextended]{svjour3}       
\smartqed  
\usepackage{cite}
\usepackage{graphicx}
\usepackage{paralist}
\usepackage{booktabs} 
\usepackage{longtable}
\usepackage{lscape}
\usepackage{appendix}
\usepackage{verbatim} 
\usepackage[colorlinks]{hyperref}
\usepackage[colorinlistoftodos]{todonotes}

 \journalname{Empirical Software Engineering}

\begin{document}
%
\title{Open Innovation Using Open Source Tools: A Case Study at Sony Mobile }


\author{Hussan~Munir \and
        Johan~Lin{\aa}ker \and  Krzysztof Wnuk \and Per Runeson \and Bj\"{o}rn Regnell   }

\institute{ H. Munir, J. Lin{\aa}ker, K. Wnuk, P. Runeson, B. Regnell  \at
           Dept of Computer Science, Lund University,  Sweden \\
            \email{[hussan.munir, johan.linaker, per.runeson, bjorn.regnell]@cs.lth.se} \\  krzysztof.wnuk@bth.se   
}

\maketitle

\begin{abstract}

\textbf{Background.} Despite growing interest of Open Innovation (OI) in Software Engineering (SE), little is known about what triggers software organizations to adopt it and how this affects SE practices. OI can be realized in numerous of ways, including Open Source Software (OSS) involvement. Outcomes from OI 
are not restricted to product innovation but also include process innovation, e.g. improved SE practices and methods.  
\textbf{Aim.} This study explores the involvement of a software organization (Sony Mobile) in OSS communities from an OI perspective and what SE practices (requirements engineering and testing) have been adapted in relation to OI. It also highlights the innovative outcomes resulting from OI.
\textbf{Method.} An exploratory embedded case study  investigates how Sony Mobile use and contribute to Jenkins and Gerrit; the two central OSS tools in their continuous integration tool chain. Quantitative analysis was performed on change log data from source code repositories in order to identify the top contributors and triangulated with the results from five semi-structured interviews to explore the nature of the commits.
\textbf{Results.} The findings of the case study include five major themes: 
\begin{inparaenum}[i)] 
\item The process of opening up towards the tool communities correlates in time with a general adoption of OSS in the organization. 
\item Assets not seen as competitive advantage nor a source of revenue are made open to OSS communities, and gradually, the organization turns more open. 
\item The requirements engineering process towards the community is informal and based on engagement. 
\item The need for systematic and automated testing is still in its infancy, but the needs are identified. 
\item The innovation outcomes included free features and maintenance, 
and were believed to increase speed and quality in development. 
\textbf{Conclusion.} 
Adopting OI was a result of a paradigm shift of moving from Windows to Linux.
This shift enabled Sony Mobile to utilize the Jenkins and Gerrit communities to make their internal development process better for its software developers and testers.
\end{inparaenum}


\end{abstract}

\section{Introduction}
\label{sec:introduction}

Software organizations have recently been exposed to new facets of openness that go beyond their experience and provide opportunities outside their organizational walls. Chesbrough~\cite{chesbrough_open_2003} explains the term \emph{Open Innovation} (OI) as \textit{``a paradigm that assumes that organizations can and should use external ideas as well as internal ideas, and internal and external paths to market, as they look to advance their technology''}. OI is based on \textit{outside-in} and \textit{inside-out} knowledge flows that help to advance technology and spark innovation. Some classical examples of inside-out are selling intellectual property while outside-in correspond to start-up acquisition and integration. There are also \textit{coupled processes}~\cite{Enkel2009} where companies give and take during co-creation by making alliances and joint-ventures. OI is fuelled by increased mobility of workers and knowledge, more capable universities, greater knowledge access and sharing capabilities that World Wide Web offers~\cite{chesbrough2014} and easier access to venture capital for start-ups. 


Open Source Software (OSS) was widely used by software organizations before the OI model became popular ~\cite{Lee2009426} and nowadays provides a common example of OI~\cite{MunirMapping15}. OSS leverages external resources and knowledge to increase innovation, product quality and to shorter time-to-market. OSS offers not only potential product innovation (e.g. by using an OSS platform of commodity parts to build differentiation parts), but potential process innovations in terms of  an implementation of new or significantly improved production or delivery  methods\cite{LinakerSurvey15}.

IBM's engagement in the Linux community in terms of patent and monetary contributions exemplifies how a firm can leverage OSS from an OI perspective. Risks and costs of development were in this case shared among other stakeholders such as Intel, Nokia, and Hitachi, which also have made significant investments in the Linux community~\cite{Lee2003}. Thanks to Linux involvement, IBM can strengthen its own business model in selling proprietary solutions for its clients running on top of Linux. Additionally, the openness of Linux also gave IBM more freedom to co-develop products with its customers~\cite{chesbrough2014}.

Software organizations that want to benefit from OI via OSS engagement need to adapt and innovate their internal software development strategies and processes. For example, influence on feature selection and road-mapping may be gained through a more active participation, as many OSS communities are based on meritocracy principles~\cite{jensen2007role}. Also, some benefits may first be fully utilized after contributing back certain parts to the OSS community~\cite{ven2008challenges}. For example, by correcting bugs, actively participating in discussions and contributing new features, a software organization might reduce maintenance cost compared to proprietary software development~\cite{stuermer_extending_2009}. Hence, in order for a firm to gain the expected benefits of products, OI process innovations may be a required step on the way forward~\cite{Lakhani2003923,Wnuk2012,Rolandsson2011576}. Existing literature does not particularly focus on how these internal SE process adaptions should be structured or executed~\cite{MunirMapping15}. Further, little is known about how OSS involvement may be utilized as an enabler and support for further innovation spread inside an organization, e.g. process, tools, or organizational innovations.

\begin{figure}[t]
\centering
\scalebox{1}
{
\includegraphics[width = \textwidth]{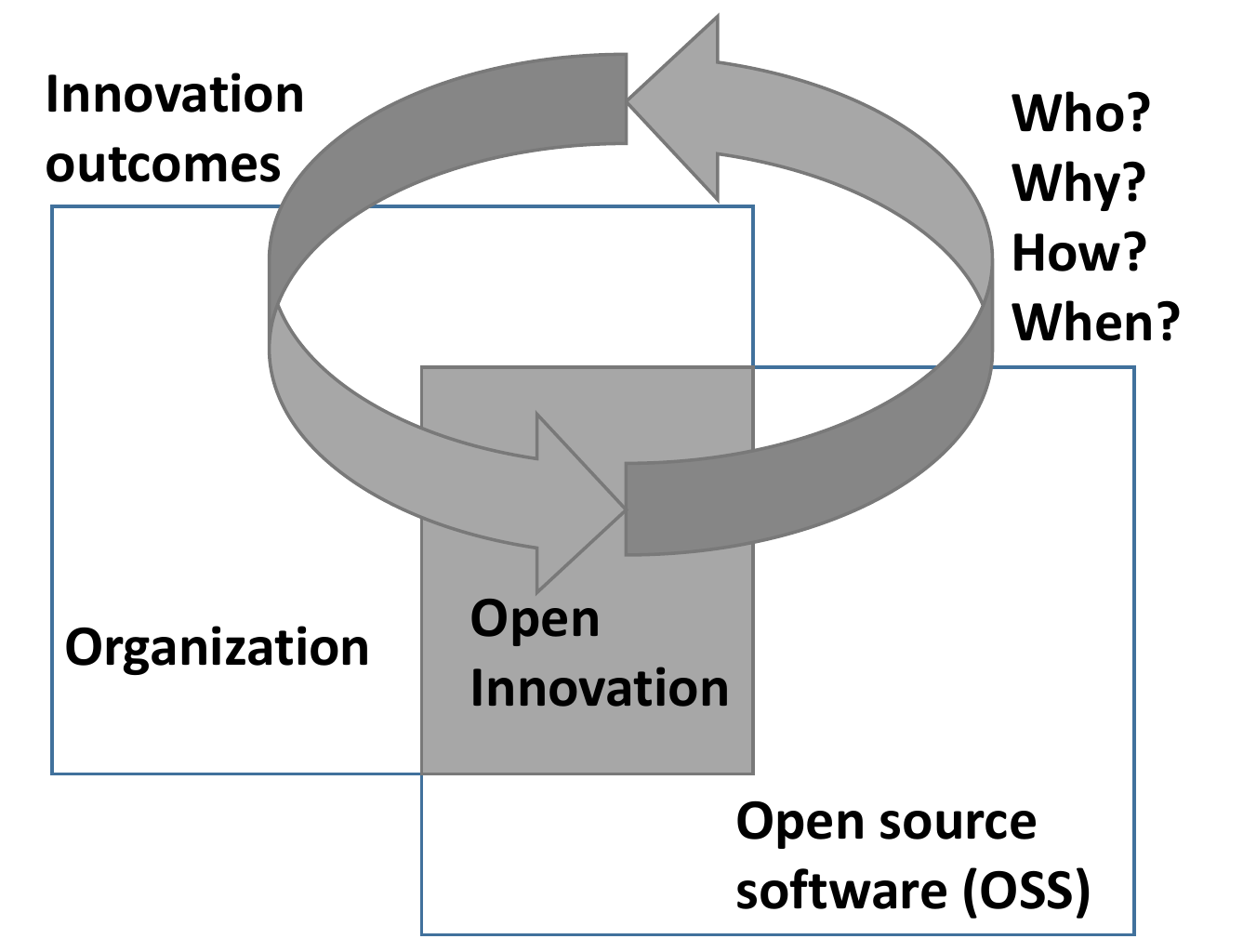}
}
\caption{Study Objectives in the intersection between proprietary organizations and open source software.}

\label{fig:Studyobjectives}
\end{figure}

In this study, we focus on identifying when, why and how a 
software organization adopts OI through the use of OSS, and what innovative outcomes can be gained (see Fig.~\ref{fig:Studyobjectives}). We investigate these aspects through a case study at Sony Mobile and how it actively participate and contribute to the communities of the two OSS tools Jenkins and Gerrit. 
These two tools are the basis of Sony Mobile's internal continuous integration tool chain. The study further investigates how external knowledge and innovation captured through the active development of these OSS tools may be transferred into the product development teams of Sony Mobile. More explicitly, this study contributes by studying how OSS may be used, not only for leveraging product innovation~\cite{LinakerSurvey15} in the tools themselves, but also how these tools can be used as enablers for process innovation in the form of improved SE practices and product quality.

This paper is structured as follows. Section \ref{sec:Relatedwork} highlights the related work and Section \ref{sec:casestudydesign} outlines the research methodology. In Sections \ref{sec:Quantitativeanalysis} and \ref{sec:Qualitativeanalysis} results from the quantitative and qualitative analysis are presented, respectively. Finally, Section \ref{sec:discussion} discussed the results, followed by conclusions in Section \ref{sec:conclusion}. 

\section{Related Work}
\label{sec:Relatedwork}
In this section, we summarize related work in OI strategies, OI challenges in SE and open source development practices inside software organizations. This section is partly based on the systematic mapping study by Munir et al.~\cite{Munir2014}.

The increased openness that OI implies poses significant challenges to software organizations in terms of securing their competitive advantage~\cite{MunirMapping15} and understanding what to contribute, when and how to maintain differentiation towards competitors that may also be involved in the OSS community~\cite{henkel_selective_2006, lindencommodification2009,Jansen2012SGO}. Related to that is the challenge of what requirements should be selected, when these should be released and how an internal roadmap should be synchronized with the OSS project's roadmap~\cite{Wnuk2012, linaaker2016firms}. These challenges highlight the need for a clear contribution strategy that software organizations should create  to focus their internal resources on value-creating activities, rather than contributing unnecessary patches or differentiating features~\cite{Wnuk2012}.

Extensive involvement in OSS communities may also bring significant challanges. Among these challenges, Daniel et al.~\cite{daniel2011} suggested that the conflict between organizational and OSS standards reduces developers' organizational commitment and it 
is strongly dependent on the degree to which developers associate themselves with organizations or OSS communities. Investing in OSS may also be costly and create differentiation and property right protection challenges, as indicated by Stuermer et al.~\cite{stuermer_extending_2009} who studied the Nokia Internet Tablet, which was based on a hybrid of OSS and proprietary software development.  


West et al.~\cite{west_creating_2008} examined the complex ecosystem surrounding Symbian Ltd. and identified three inherent difficulties for organizations leading an OI ecosystem: 1) prioritizing the conflicting needs of heterogeneous ecosystem participants, 2) knowing the ecosystem requirements for a product that has yet to be created, and 3) balancing the interests of those participants against those of the ecosystem leader.

Looking at OI strategies, Dahlander \& Magnusson~\cite{Dahlander2008} show how organizations may access OSS  communities in order to extend the firm's resource base, align the organization's strategy with that of the OSS community, and/or assimilate the community in order to integrate and share results with them. The same authors explained that depending on how open a firm chooses to be in regards to their business model, different strategies may be enforced, e.g. symbiotically giving back result to the community, or as a free-rider keeping modifications and new functionality to oneself~\cite{Dahlander2005481}. Some strategies include:
\begin{itemize}
    \item selectively revealing - differentiating parts are kept internal while commodity parts are made open~\cite{henkel_selective_2006, West20031259}. This requires continuous assessment of what parts are to be considered commodity as opposed to differentiating value.
    \item licensing schemas (cf. Dual-licensing~\cite{chesbrough2007open}), technology may be fully disclosed, but under 
    a restrictive license~\cite{West20031259}. Alternatively, everything may be disclosed under open and transparent conditions~\cite{chesbrough2007open}.
\end{itemize}


Henkel~\cite{henkel_selective_2006} reports how small organizations reveal more, as they are likely to benefit from the external development support. Component manufacturers also reported to contribute a lot as they have a good protection of the hardware they sell; software is seen as a complementary asset. In a follow-up study, Henkel~\cite{henkel_emergence_2013} further reported how openness had become a competitive edge, as customers had started to request even more revealing.

Dahlander \& Wallin~\cite{Dahlander20061243} show how having an employee in the community can be an enabler for the organizations to not only gain a good reputation but also to influence the direction of the development towards the organizations' own interests. However, to gain the roles needed to commit or review code written by community developers, individuals need to contribute and become an active part of the communities as these are often based on the principles of meritocracy~\cite{jensen2007role}.



Inner Source~\cite{stol2014key} has gained interest among researchers and practitioners as a way to adapt OSS practices at software organizations. Such hybrids of commercial and OSS practices~\cite{mockuswhy2002} could include using the OSS style project structure, where a core team of recognized experts has the power to commit code to an official release, and a much larger group contributes voluntarily in many ways.

\textbf{Summary}. 
Research has shown a lot of interest for OI and its different applications~\cite{West2013}, including leveraging OSS for OI ~\cite{MunirMapping15}. However, the focus is mostly limited to management and strategic aspects, e.g., ~\cite{Dahlander2008, west2013evolving, stuermer_extending_2009}, with some exception of inner sourcing~\cite{Morgan2012, stol2014key}. Little is still known about what triggers software organizations to adopt OSS from an OI perspective and how this affects SE practices~\cite{MunirMapping15}. 

This paper adds to existing knowledge by focusing on the use of OSS from an OI perspective in an organization that seek to complement its internal product development and process innovation~\cite{LinakerSurvey15} with the use of external knowledge from OSS communities.  
Furthermore, this study aims to improve 
our understanding of what and how a software organization can open up and how SE practices are adapted to deal with the openness to OSS communities.

\section{Case Study Design}
\label{sec:casestudydesign}
Below we describe the research design of this study. We explain the research questions, the structure of the case study design, and the methodologies used for data collection as well as for the quantitative and qualitative analysis.

\subsection{Research Questions}
\label{sec:researchquestions}

\begin{table*}[t]
  \centering
  \caption{Research questions with description}
    \begin{tabular}{p{5cm} p{5cm}}
    \textbf{Research Questions} &
      \textbf{Objective}
      \\
      \toprule
    \textbf{RQ1:} How and to what extent is Sony Mobile involved in the communities of Jenkins and Gerrit? &
      To characterize Sony Mobile's involvement and identify potential interviewees.
      
      \\
      \midrule
    \textbf{RQ2:} What is the motivation for Sony Mobile to adopt OI?   &
      To explore the transition from a closed innovation process to an OI process.
      \\
      \midrule
    \textbf{RQ3:} How does Sony Mobile take a decision to make a project or feature open source? &
      To investigate what factors affect the decision process when determining whether or not Sony Mobile should contribute functionality. 
            \\
      \midrule
   \textbf{RQ4:} What are the innovation outcomes as a result of OI participation? &  To explore the vested interest of Sony Mobile as they moved from a closed innovation model to an OI model.    
      \\
      \midrule
    \textbf{RQ5:} How do the requirements engineering and testing processes interplay with the OI adoption? &
      To investigate the requirements engineering and testing processes and how they deal with the special complexities and challenges involved due to OI.
      
      \\
      \bottomrule
    \end{tabular}%
  \label{tab:Researchquestions}%
\end{table*}%



The focus of this study is on how software organizations use OSS projects from an OI perspective, what triggers them to open up  and how this impacts the organizations' innovative performance and their SE practices (see Fig.~\ref{fig:Context}). We investigate these aspects through a case study at Sony Mobile, and how they actively participate and contribute to the communities of the two OSS tools Jenkins~\cite{Jenkins} and Gerrit~\cite{Gerrit}. Both tools constitute pivotal parts in Sony Mobile's internal continuous integration tool chain. 

The study further investigates how external knowledge and innovation captured through the development of these OSS tools, may be transferred into the product development teams of Sony Mobile. More explicitly, this study contributes by studying how OSS may be used, not only for leveraging product innovation~\cite{LinakerSurvey15} in the tools themselves, but also how these tools can be used as enablers for process innovation in the form of improved SE practices and tools within the organization.

\begin{enumerate}
\item \textbf{\textit{Jenkins}} is an open source build server that runs on a standard servlet container e.g. Apache Tomcat. It can handle Maven and Ant instructions, as well as execute custom batch and bash scripts. It was forked from the Hudson build server in 2010 due to a dispute between Oracle and the rest of the community.
\item \textit{\textbf{Gerrit code review}} is an OSS code review tool created by Google in connection with the Android project in 2007. It is tightly integrated with the software configuration management tool GIT, working as a gatekeeper, i.e. a commit needs to be reviewed and verified before it is allowed to be merged into the main branch.
\end{enumerate}

Based on this background, and the research gap identified in earlier work~\cite{MunirMapping15}, we formulate our research questions to study the OI in Sony Mobile in an exploratory manner (see Table~\ref{tab:Researchquestions}). \textit{RQ1} addresses the extent to which Sony Mobile is involved in the Jenkins and Gerrit communities and its key contribution areas (i.e. bug fixes, new features, documentation etc.). \textit{RQ2} and \textit{RQ3} explore the rationale behind Sony Mobile's transition from closed innovation to OI. \textit{RQ4} highlights the key innovation outcomes realized as a result of 
openness. Finally, \textit{RQ5} aims at understanding whether or not the existing requirements engineering and testing processes have the capacity to deal with the OI challenges in SE. \textit{RQ1} is answered with the help of quantitative analysis of repository data, while the remaining four research questions (\textit{RQ2, RQ3, RQ4, RQ5}) are investigated using qualitative analysis of interview data.  

%
%

\subsection{Case Selection and Units of Analysis}
\label{sec:unitsofanalysis}	
Sony Mobile is a multinational corporation with roughly 5,000 employees, developing embedded devices. The studied branch focuses on developing Android-based phones and tablets and has 1600 employees, of which 900 are directly involved in software development. Sony Mobile develops software in an agile fashion and applies software product line management with a database of more than 20,000 features suggested or implemented across all product lines~\cite{Pohl2005}. 

\begin{figure}[t]
\centering
\scalebox{1}
{
\includegraphics[clip, trim = {0 0 0 3cm},width=\textwidth]{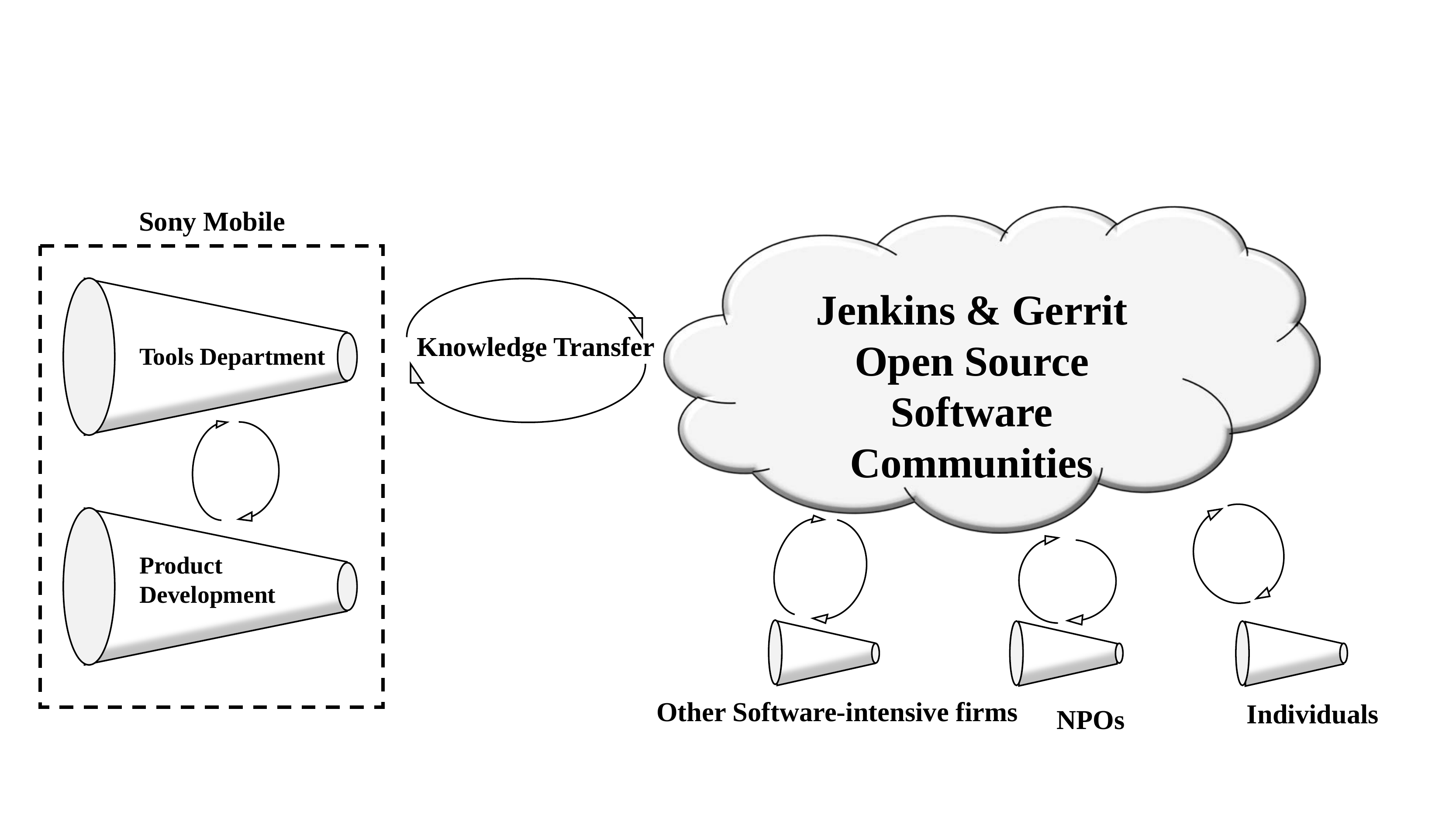}
} 
\caption{The Jenkins and Gerrit OSS communities surrounded by Sony Mobile and other members. Arrows represent knowledge transfer in and out of the community members such as other software organizations, non profit organizations (NPO) and individuals, which in turn are illustrated by funnels, commonly used in OI literature~\cite{chesbrough_open_2003}.}
\label{fig:Context}
\end{figure}

However, in order to work with OSS communities, namely Jenkins and Gerrit Sony Mobile created a designated tools department to acquire and integrate the external knowledge to improve the internal continuous integration process. The continuous integration tool chain used by Sony Mobile is developed, maintained and supported by an internal tools department. The teams working on phones and tablets are thereby relieved of this technical overhead. During the recent years, Sony Mobile has transitioned from passive usage of the Android codebase into active involvement and community contribution with many code commits to Jenkins and Gerrit. This maturity resulted in a transition from closed innovation to OI ~\cite{chesbrough_open_2003}, assuming that business values are created or captured as an effect.

From an OI perspective, there are interactions between the Tools department and the Jenkins and Gerrit communities (see Fig.~\ref{fig:Context}). The in- and outgoing transactions, visualized by the arrows in Fig.~\ref{fig:Context}, are data and information flows, e.g. ideas, support and commits, can be termed as a coupled innovation process~\cite{Enkel2009}. The exchange is continuous and bi-directional, and brings product innovation into the Tools department in the form of new features and bug fixes to Jenkins and Gerrit. 

The Tools department can, in turn, be seen as a gate between external knowledge and the other parts of Sony Mobile (see Fig. \ref{fig:Context}). The Tools department accesses, adapts and integrates the externally obtained knowledge from the Jenkins and Gerrit communities into the product development teams of Sony Mobile. This creates additional transactions inside Sony Mobile which can be labeled as process innovation~\cite{osloManual2005} in the sense that new tools and ways of working improve development efficiency and quality. This relates to the internal complementary assets need that is mentioned as an area for future research by Chesbrough et al.~\cite{chesbrough2006}.

We conducted a case study design with Jenkins and Gerrit as units of analysis~\cite{Runesoncasestudy2012} as these are the products in which the exchange of data and information enable further innovation inside Sony Mobile.


\subsection{Case Study Procedure}
\label{sec:CaseStudyProcedure}
We performed the following steps. 
\begin{enumerate}
\item Preliminary investigation of Jenkins and Gerrit repositories.
\item Mine the identified project repositories.
\item Extract the change log data from the source code repositories. 
\item Analyze the change log data (i.e. stakeholders, commits etc).
\item Summarize the findings from the change log data to answer \textit{RQ1}.
\item Prepare and conduct semi-structured interviews to answer \textit{RQ2--RQ5}.

\item Synthesize data.  
\item Answer the research questions \textit{RQ1--RQ5}.
\end{enumerate}


\subsection{Methods for Quantitative Analysis}
\label{sec:MethodsQuantitativeAnalysis}
To understand Sony Mobile's involvement in the OSS tools (\textit{RQ1}), we conducted quantitative analysis of commit data in the source code repositories of Jenkins and Gerrit.

\subsubsection{Preliminary Investigation of Jenkins and Gerrit Commits}
\label{sec:PreliminaryInvestigationjenkinsgerrit}


A \textit{commit} is a snapshot of a developer's files after reaching a code base state. The number of lines of code in a commit may vary depending upon the nature of the commit (e.g. new implementation, update etc.)~\cite{hattorinature2008}. The comment of a commit refers to a textual message related to the activity that generates the updated new piece of code. It ranges from a simple note to a
detailed description, depending on the project's conventions. In this study, we used the keywords provided by Hattori \cite{hattorinature2008} in his study as a reference point to classify the commit messages (see Table \ref{tab:commitsclassfication}). 

We mined the source code repositories of Jenkins and Gerrit to extract the commit id, date, committer name, committer email and commit description message for each commit, with the help of the tool CVSAnlY~\cite{cvsanaly}. The extracted data was stored locally in a relational database with a standard data scheme, independent of the analyzed code repository. The structure of the database allows a quantitative analysis to be done by writing SQL queries. 
The number of commits per committer were added together with the name and email of the committer as keys. 

We extracted the affiliations of the committers from their email addresses by filtering them on the domain, e.g., john.doe@sonymobile.com was classified with a Sony Mobile affiliation. It is recognized that committers may not use their corporate email addresses when contributing their work, since parts of their work could be contributed privately or under the umbrella of other organizations than their employer. To triangulate and complement this approach, a number of additional sources were used, as suggested by earlier research~\cite{bird2012examining,gonzalez2013understanding}. First, social media sites as LinkedIn, Twitter and Facebook were queried with keywords from the committer, such as the name, variations of the username and e-mail domain. Second, unstructured sources such as blogs, community communication (e.g., comment-history, mailing-lists, IRC logs), web articles and firm websites were consulted. 


Sony Mobile turned out to be one of the main organizational affiliations among the committers to Gerrit while no evidence of commits to the Jenkins core community was identified. The reason for this was that Jenkins is a plug-in-based community, i.e. there is a core component surrounded by approximately 1,000 plug-ins of which each has a separate source code repository and community. Our initial screening had only covered the core Jenkins component. After analyzing forum postings, blog posts and reviewing previously identified committers, a set of Jenkins plug-ins, as well as two Gerrit plug-ins, were identified, which then were also included in our analysis. The following Open Source projects were included for further analysis:

\begin{itemize}
\item Gerrit\footnote{https://www.openhub.net/p/gerrit}
\item PyGerrit (Gerrit plug-in)\footnote{https://www.openhub.net/p/pygerrit}
\item Gerrit-events (Gerrit plug-in)\footnote{https://www.openhub.net/p/gerrit-events}
\item Gerrit-trigger (Jenkins plug-in)\footnote{https://github.com/jenkinsci/gerrit-trigger-plugin}
\item Build-failure-analyzer (Jenkins plug-in)\footnote{https://www.openhub.net/p/build-failure-analyzer-plugin}
\item External-resource-viewer (Jenkins plug-in)\footnote{https://github.com/jenkinsci/external-resource-dispatcher-plugin}
\item Team-views (Jenkins plug-in)\footnote{https://github.com/jenkinsci/team-views-plugin}
\end{itemize}
\label{sec:methodquantitativeanalysis}

\subsubsection{Classification of Commit Messages}
\label{sec:CommitsClass}
Further analysis included creating the list of top committers combined with their yearly activity (number of commits) in order to see how Sony Mobile's involvement evolved over time. Next, we characterized and classified the commits made by Sony Mobile to the corresponding communities by following the criteria defined by Hattori et al.~\cite{hattorinature2008}. This was done manually by analyzing the description messages of the commits and searching for keywords (see Table \ref{tab:commitsclassfication}), and then classifying the commits in one of the following categories:

\begin{table}[t]
  \centering
  \caption{Keywords used to classify commits taken from Hattori~\cite{hattorinature2008}.}
    \begin{tabular}{p{2.2cm} p{2.2cm} p{2.2cm} p{2.3cm} p{2.3cm}}
    \toprule
    \textbf{Forward Engineering} & \textbf{Re-engineering} & \textbf{Corrective Engineering} & \textbf{Management} \\
    \midrule
    IMPLEMENT & OPTIMIZ & BUG   & CLEAN \\
    ADD   & ADJUST & ISSUE & LICENSE \\
    REQUEST & UPDATE & ERROR & MERGE \\
    NEW   & DELET & CORRECT & RELEASE \\
    TEST  & REMOV & PROPER & STRUCTURE \\
    START & CHANG & DEPRAC & INTEGRAT \\
    INCLUD & REFACTOR & BROKE & COPYRIGHT \\
    INITIAL & REPLAC &       & DOCUMENTATION \\
    INTRODUC & MODIF &       & MANUAL \\
    CREAT & ENHANCE &       & JAVADOC \\
    INCREAS & IMPROV &       & COMMENT \\
          & DESIGN CHANGE &       & MIGRAT \\
          & RENAM &       & REPOSITORY \\
          & ELIMINAT &       & CODE REVIEW \\
          & DEUPLICAT &       & POLISH \\
          & RESTRUCTUR &       & UPGRADE \\
          & SIMPLIF &       & STYLE \\
          & OBSOLETE &       & FORMATTING \\
          & REARRANG &       & ORGANIZ \\
          & MISS  &       & TODO \\
          & ENHANCE &       &  \\
          & IMPROV &       &  \\
    \bottomrule
    \end{tabular}%
  \label{tab:commitsclassfication}%
\end{table}%


\textbf{Forward engineering}
activities refer to the incorporation of new features and implementation of new requirements including the writing new test cases to verify the requirements. 
\textbf{Re-engineering}
activities deal with re-factoring, redesign and other actions to enhance the quality of the code without adding new features. 
\textbf{Corrective engineering}
activities refer to fixing defects in the software. 
\textbf{Management activities} are related to code formatting, configuration management, cleaning up code and updating the documentation of the project.

Multiple researchers were involved in the commit message classification process. After defining the classification categories, Kappa analysis was performed to calculate the inter-rater agreement level. First, a random sample of 34\% of the total commit messages were taken to classify the commit messages and Kappa was calculated to be 0.29. Consequently, disagreement was discussed and resolved since the inter-rater agreement level was below substantial agreement range. Afterwards, Kappa was calculated again and found to be 0.94.

\subsection{Methods for Qualitative Analysis}
The quantitative analysis had laid a foundation to understand the relation between Sony Mobile, and the Jenkins and Gerrit communities. Therefore, in the next step we added a qualitative view by interviewing relevant people inside Sony Mobile in order to address \textit{RQ2--RQ5}. Interview questions are listed in the Appendix. 

\subsubsection{Interviewee Selection}
The selection of interviewees was based on the committers identified in the initial screening of the projects. Three candidates were identified and contacted by e-mail (Interviewees 1, 2 and 3, see Table~\ref{tab:interviewees}). Interviewees 4 and 5 were proposed during the initial three interviews. The first three are top committers to the Jenkins and Gerrit communities, giving the view of Sony Mobile's active participation and involvement with the communities. It should be noted that interviewee I3, when he was contacted, had just left Sony Mobile for a smaller organization dedicated to Jenkins development. His responsibilities as the tools manager for Jenkins at Sony Mobile were taken over by interviewee I4. Interviewee I4 is a Software Architect in the Tools department involved further down in Sony Mobile's continuous integration tool chain and gives an alternative perspective on the OSS involvement of the Tools department as well as a higher, more architectural view on the tools. Interviewee I5 is an upper-level manager responsible for Sony Mobile's overall OSS strategy, which could contribute with a top-down 
perspective to the qualitative analysis.


\begin{table*}[t]
  \centering
  \caption{Interviewee demographics.}
  \scalebox{1.0}{
    \begin{tabular}{p{2cm} p{.5cm} p{2.5cm}  p{2cm}   p{2.5cm}}
    \toprule
    \textbf{Anonymous name } &
    ID&
      \textbf{Tools involvement} &
      \textbf{Years of experience} &
      \textbf{Role}
      \\
      \midrule
    Interviewee 1 &
    I1
    &
      Jenkins &
      8 &
      Tools manager for Jenkins
      \\
      \midrule
    Interviewee 2 &
    I2&
      Jenkins and Gerrit &
      6 &
      Team lead, Tools manager for Gerrit
      \\
      \midrule
    Interviewee 3 &
      I3&
      Jenkins &
      7 &
      Former tools manager Jenkins
      \\
      \midrule
    Interviewee 4 &
      I4&
      Second line after Jenkins and Gerrit Build artifacts and channel distribution &
      8 &
      Software Architect
      \\
      \midrule
    Interviewee 5 &
      I5&
      Open Source policy in general &
      20+ &
      Upper-level manager responsible for overall Open Source strategy
      \\
      \bottomrule
    \end{tabular}%
    }
  \label{tab:interviewees}%
\end{table*}%

The interviews were semi-structured, meaning that interview questions were developed in advance and used as a frame for the interviews, but still allowing the interviewers to explore  other relevant 
findings during the interview wherever needed. The two first authors were present during all five interviews, with the addition of the third author during the first and fifth ones. Each interviewer took turns asking questions, whilst the others observed and took notes. Each interview was recorded and transcribed. A summary was also compiled and sent back to the interviewees for a review. Any misunderstandings or corrections could then be sorted out. The duration of the 
interviews varied from 45 to 50 minutes.

\subsection{Validity Threats}
\label{sec:Validity Threats}
This section highlights the validity threats related to the case study. Four types of validity threats~\cite{Runesoncasestudy2012} are addressed with their mitigation strategies.

\subsubsection{Internal Validity}
This \label{sec:internalValidity} concerns causal relationships and the introduction of potential confounding factors. 

\textit{Confounding factors}. To mitigate the risk of introducing confounding factors, the study was performed on the tools level instead of an organizational level to ensure that the innovation outcomes are merely the result of adopting OI. Performing the study on an organization level introduces the risk of confounding the innovation outcomes as a result of a product promotion or financial investment etc. instead of the use of external knowledge from OSS communities. Therefore, a more fine-grained analysis on the OSS tools level was chosen to minimize the threat of introducing confounding factors. 

\textit{Subjectivity.} It was found in the study that Sony Mobile does not use any general innovation metrics to measure the impact of OI. 
Therefore, researchers 
had to rely on qualitative data. 
This leads to the risk of introducing subjectivity while inferring innovation outcomes as a result of OI adoption. In order to minimize this risk, the first two authors independently performed the analysis and the remaining authors reviewed it to make the synthesis more objective. Moreover, findings were sent back to interviewees for validation. Furthermore, subjectivity was minimized by applying the commit messages classification criteria proposed by Hattori et al.~\cite{hattorinature2008}. During the analysis, the disagreements were identified using Kappa analysis and resolved to achieve a substantial agreement. 

\textit{Triangulation.} In order to mitigate the risk of identifying the wrong innovation outcomes, we used multiple data sources by mining the Jenkins and Gerrit source code repositories prior to conducting interviews. Furthermore, we also performed observer triangulation during the whole course of the study to mitigate the risk of introducing researcher bias.

\subsubsection{External Validity}
\label{sec:ExternalValidity}
This refers to the extent it is possible to generalize the study findings to other contexts. The scope of this study is limited to a software organization utilizing the notion of OI to accelerate its innovation process. The selected case organization is a large-scale organization with an intense focus on software development for embedded devices. Moreover, Sony Mobile is a direct competitor of all the main stream organizations making Android phones. The involvements by other stakeholders in the units of analysis (Jenkins and Gerrit) indicate their adoption of Google's tool chain to improve their continuous integration process. Therefore, the findings of this study may be generalized to major stakeholders identified for their commits to Jenkins and Gerrit, and other OSS tools used in the tool chain development. 
Our findings may also be relevant to software organizations with similar context, domain and size as Sony Mobile.

\subsubsection{Construct Validity}
\label{sec:ConstructValidity}

This refers to what extent the operational measures that are studied really represent what researcher has in mind, and what is investigated according to the research questions~\cite{Runesoncasestudy2012}. 
We took the following actions 
to minimize construct validity threats.

\textit{Selection of interviewees}. We conducted a preliminary quantitative analysis of the Jenkins and Gerrit repositories to identify the top committers and to select the relevant interviewees. The selection was performed based on the individuals' commits to Jenkins or Gerrit. Moreover, recommendations were taken from interviewees for suitable further candidates to attain the required information on OI. Process knowledge, role, and visible presence in the community were the key selection factors.

\textit{Reactive bias}. Researchers presence might limit or influence the interviewees and causing them to hide facts or respond after assumed expectations. This threat was limited by the presence of a researcher that has a long research collaboration record with Sony Mobile and explained confidentiality rules. Furthermore, interviewees were ensured anonymity both within the organization and externally in the OSS community. 

\textit{Design of the interviews}. All authors validated the interview questionnaire followed by a pilot interview with an OSS Jenkins community member in order to avoid misinterpretation of the interview questions. 

\subsubsection{Reliability}
\label{sec:Conclusion Validity}
The reliability deals with to what extent the data and the analysis are dependent on the specific researcher, and the ability to replicate the study. 

\textit{Member checking.} To mitigate this risk, multiple researchers individually transcribed and analyzed the interviews to make the findings more reliable. In addition, multiple data sources (qualitative and quantitative) were considered to ensure the correctness of the findings and cross-validate them. All interviews were recorded, transcribed and sent back to interviewees for validation. The commit database analysis was performed and validated by multiple researchers. 

\textit{Audit trail.} Researchers kept track of all the mined data from OSS code repositories as well as interview transcripts in a systematic way to go back for validation if required. Finally, this study was not ordered by Sony Mobile to bring supporting evidence for OI adoption. Instead the idea was to keep the study design and findings as transparent as possible without making any adjustments in the data except for the anonymizing the interviewees. The results were shared with Sony Mobile prior to submitting the study for publication.

\section{Quantitative Analysis}
\label{sec:Quantitativeanalysis}
This section presents a quantitative analysis of commits made to eight OSS projects, namely: Gerrit, pyGerrit, Gerrit-events, Gerrit-trigger, Build-failure-analyzer, External-resource-viewer and Team-views as depicted in section \ref{sec:PreliminaryInvestigationjenkinsgerrit}. It should be noted that the seven latter projects are plugins to Gerrit and Jenkins, i.e., not part of the core projects. In the analysis we investigated the types of commits made (see Section~\ref{sec:CommitsClass}), and in what proportion these were made by Sony Mobile over time, as well as compared to other major organizations.

\subsection{Gerrit}
\label{sec:Gerrit}

\begin{table}[t]
    \center
    \begin{tabular}{lc|c|c|c|c|c|c}
         \textbf{Commits classification} & \textbf{2010} & \textbf{2011} & \textbf{2012} & \textbf{2013} & \textbf{2014} & \textbf{Total}   \\
         \toprule
         Forward Engineering &  65 & 44 & 264 & 373 & 207 & 953 \\ \midrule
         Re-engineering & 38 & 65 & 240 & 336 & 190  & 869\\ \midrule
         Corrective engineering & 10 & 12 & 59 & 62 & 26 & 169\\ \midrule
         Management & 12 & 15 & 96 & 171 & 73 & 367\\  \bottomrule
        
         \textbf{Total} & 125 & 136 & 659 & 942 & 496 & 2358\\ 
    \end{tabular}
    \caption{Sony Mobile's commits to Gerrit analyzed per year.}
    \label{tab:GerritYearByYear}
\end{table}

The two largest categories of commits for Gerrit are forward engineering (953 commits) and re-engineering (869 commits), followed by management commits (367 commits) and corrective engineering commits (169 commits), see Table~\ref{fig:bubbleplot}. This dominance of forward and re-engineering commits remained stable between 2010 and 2014, see Table~\ref{tab:GerritYearByYear}. Sony Mobile presented the first Android-based mobile phone in March 2010 and as can be seen from the analysis also became active in contributions to Gerrit with a total of 125 contributions in 2010. From 2012 the number of forward and re-engineering commits became more equal each year suggesting that Sony Mobile was not only contributing new features but also actively helping in increasing the quality of the current features and re-factoring.


\begin{table}[htbp]
  \centering
  \caption{Classification of Sony Mobile's commits to OSS tools based on hattori's criteria~\cite{hattorinature2008}}
    \begin{tabular}{p{2cm} p{2cm} p{2cm} p{2cm} p{2cm}}
    \toprule
    \textbf{Tools} & \textbf{Forward Engineering} & \textbf{Re-Engineering} & \textbf{Corrective Engineering} & \textbf{Management} \\
    \midrule
    Gerrit &  953 & 869   & 169   & 367 \\ \midrule
    pyGerrit & 27    & 18    & 19    & 36 \\ \midrule
    Gerrit-events & 27    & 18    & 19    & 36 \\ \midrule
    Gerrit-trigger & 60    & 40    & 76    & 135 \\ \midrule
    Build-failure-analyzer & 60    & 19    & 17    & 36 \\ \midrule
    External-resource-viewer & 28    & 8     & 8     & 6 \\ \midrule
    Team-views & 7     & 0     & 0     & 5 \\
    \bottomrule
    \end{tabular}%
  \label{fig:bubbleplot}%
\end{table}%

\begin{figure*}[hbtp]
\centering
\includegraphics[width=\textwidth]{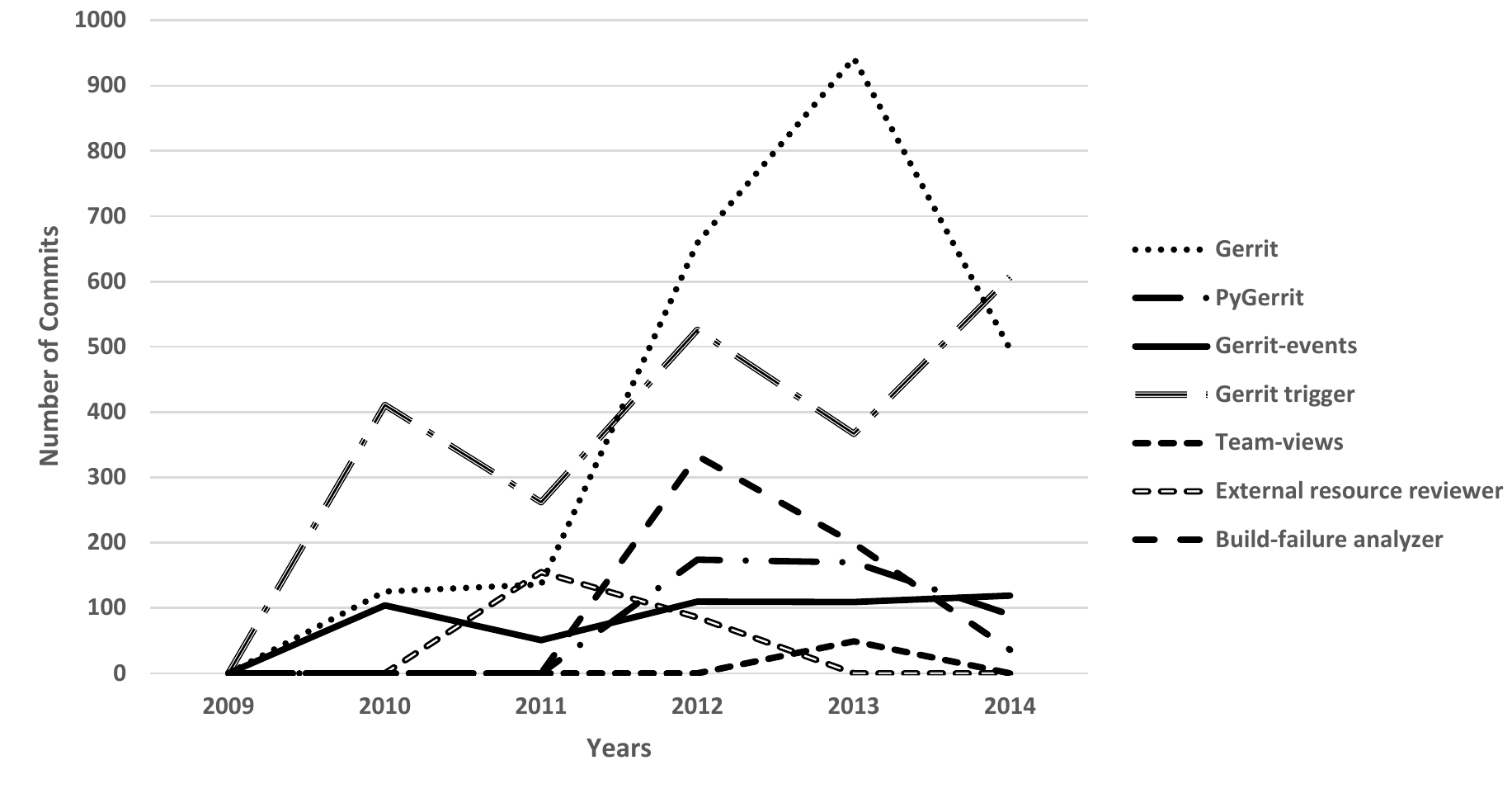}
\caption{Sony Mobile's commits for all OSS tools per year}
\label{fig:allPlugin}
\end{figure*}

The number of forward engineering and re-engineering commits remained high and we notice a substantial decrease of corrective engineering and management commits. The decrease of management commits may suggest that Sony Mobile reached a high level of compatibility of its code review processes and therefore requires fewer commits in this area. This data shows an interesting pattern in joining an OSS community. Since Sony Mobile is a large organization with several complex processes, their joining of the Gerrit community had to be associated with a substantial number of forward engineering and re-engineering commits. This entry to the community lowered the transition time and enabled faster synchronization of the code review processes between the Android community players and Sony Mobile. At the same time, Sony Mobile contributed several substantial features from the first year of participation which is positive for the community. Figure~\ref{fig:allPlugin} shows the progression of commits made by Sony Mobile to all OSS tools between year 2009 and 2014.

\begin{table}[htbp]
  \centering
  \caption{Percentage of Sony Mobile's contribution compared to other Software organizations}
    \begin{tabular}{p{3cm} p{.7cm} p{.7cm} p{.9cm} p{.7cm} p{.7cm} p{.7cm} p{.7cm}}
    \toprule
    \textbf{Tools} & \textbf{Sony} & \textbf{Google} & \textbf{Ericsson} & \textbf{HP} & \textbf{SAP} & \textbf{Intel} & \textbf{Others} \\
    \midrule
    Gerrit & 8.2  & 38.5 & 0     & 0     & 10.7 & 0     & 42.5 \\ \midrule
    PyGerrit & 97.5  & 0     & 0     & 0     & 0     & 0     & 2.4 \\ \midrule
    Gerrit-event & 66.1  & 0     & 3.3 & 4.1 & 0.2 & 2 & 24.2 \\ \midrule
    Gerrit-trigger & 65.2  & 0     & 9.1 & 2.4 & 0.7 & 1.3 & 21.2 \\ \midrule
    Team-views & 100   & 0     & 0     & 0     & 0     & 0     & 0 \\ \midrule
    External-resource-reviewer & 89.6  & 1.5 & 4.8 & 0     & 0     & 0     & 4.1 \\ \midrule
    Build-failure-analyzer & 85.5  & 0     & 0     & 0     & 0     & 0     & 14.4 \\
    \bottomrule
    \end{tabular}%
  \label{tab:Sonymobilecontributionpercentages}%
\end{table}%

\subsubsection{PyGerrit}
\label{sec:PyGerrit}
PyGerrit is a Python library that provides a way for clients to interact with Gerrit. %
As can be seen in Table \ref{tab:Sonymobilecontributionpercentages}, Sony Mobile initiated this plug-in and is the biggest committer to it, representing 97.5\% of the commits. Management commits are the most frequent category, followed by forward engineering commits. This suggests that some code formatting changes, cleaning up code and documentation commits were delivered by Sony Mobile after opening up this plug-in to the community. Sony Mobile's yearly contribution analysis shows a steady growth since its introduction in 2011 (see Fig.~\ref{fig:allPlugin}). 

\textbf{Conclusion:} This indicates that companies that want the communities to accept their plug-ins should be prepared to dedicate effort on management type of commits to increase the code's quality and documentation and therefore enable other players to contribute.


\subsubsection{Gerrit-event}
\label{sec:Gerrit-Event}
Gerrit-event is a Java library used primarily to listen to stream-events from Gerrit Code Review and to send reviews via the SSH CLI or the REST API. It was originally a module in the Jenkins Gerrit-trigger plug-in and is now broken out to be used in other tools without the dependency to Jenkins. Table \ref{tab:Sonymobilecontributionpercentages} shows that apart from Sony Mobile(66.1\%), HP(4.1\%), SAP(0.2\%), Ericsson(3.3\%) and Intel(2\%) commits reveal that they are also using Gerrit-event in their continuous integration process. Sony Mobile started contributing to Gerrit-event in 2009 and since then seem to be the largest committer along with its competitors (see Table \ref{tab:Sonymobilecontributionpercentages}). Similarly, to the PyGerrit plug-in, management and forward engineering commits dominate and Sony Mobile is the main driver of features to this community.   

\textbf{Conclusion:} Sony Mobile turns out to be the biggest contributor in Gerrit-event where the focus is mostly on adding new features (forward engineering) based on the internal organizational needs.




\subsection{Jenkins}
\label{sec:Jenkins}

Commits from Sony Mobile to Jenkins could not be identified in the core product but to a various set of plug-ins (see Table \ref{tab:Sonymobilecontributionpercentages}). The ones identified are:
\begin{itemize}
\item Gerrit-trigger
\item Build-failure-analyzer
\item External resource-reviewer
\item Team-views
\end{itemize}

\subsubsection{Gerrit-trigger}
\label{sec:GerritTrigger}

This plug-in triggers builds on events from the Gerrit code review system by retrieving events from the Gerrit command stream-events, so the trigger is pushed from Gerrit instead of pulled as scm-triggers usually are. Multiple builds can be triggered by one change-event, and one consolidated report is sent back to Gerrit. This plug-in (see Table \ref{tab:Sonymobilecontributionpercentages}) seems to attract the most number of commits with the percentage of 65.2\% from Sony Mobile. 135 commits were classified as management and 76 as corrective engineering. In this case, the majority of the commits were not forward or re-engineering, which may suggest that Sony Mobile was more interested in increasing the code quality and fixing the bugs rather than extending it. It seems logical as for the Jenkins community new functionality can be realized in the form of a new plug-in rather than extending the current plug-ins.

\textbf{Conclusion:} Adding plug-ins allows greater flexibility but increases the total number of parallel projects to manage and maintain by the community.



\subsubsection{Build-failure-analyzer}
\label{sec:Buildfailureanalyzer}
This plug-in scans build logs and other files in the workspace for recognized patterns of known causes to build failures and displays them on the build page for quicker recognition of why the build failed. As can be seen in see Table \ref{tab:Sonymobilecontributionpercentages}, Sony Mobile came out as the largest committer (85.5\%) to the Build-failure-analyzer. One possible explanation for the lack of contribution from the other software organizations is that this plug-in might be very specific to the needs of Sony Mobile, but they made it open to see if the community shows interest in contributing to further development efforts.  

 Forward engineering and management commits are the two most common categories. Moreover, the number of commits have declined after 2012 and Table \ref{fig:bubbleplot} shows a relatively  low numbers of corrective engineering (17) and re-engineering (19) commits, which seem to indicate the maturity of this plug-in in terms of quality and functionality. 
 
 \textbf{Conclusion:} We hypothesize that after creating and contributing the core functionality for a given plug-in, the number of forward commits declines and further advances are realized in a form of a new plug-in.

\subsubsection{External-resource-viewer}
\label{sec:Externalresourceviewer}
This plug-in adds support for external resources in Jenkins. An external resource is something
attached to a Jenkins slave and can be locked by a build, to get exclusive access to it, then released after the build is done. Examples of external resources are phones, printers and USB devices. Like Build-failure-analyzer, Sony Mobile's is the top commiter with the largest contribution percentage of 89.6\% compared to Google (1.48\%) and Ericsson (4.8\%). Moreover, the majority of the commits are classified as forward engineering, suggesting that Sony Mobile has come up with the majority of the functionality to this plug-in. As the number of corrective engineering and re-engineering commits remained low (8 commits in each category), we can assume that the contributed code was high quality. 

\textbf{Conclusion:} This data suggest a hypothesis that companies that frequently interact with OSS communities learn to contribute high quality code and possibly keep the same quality standards for other development initiatives.

\subsubsection{Team-views}
\label{sec:Teamviews}
This plug-in provides teams, sharing one Jenkins master, to have their own area with team-specific views. 
Sony Mobile turned out to be the only committer for this tool (see Table \ref{tab:Sonymobilecontributionpercentages}), which implies that Team-views is tailored for the needs of Sony Mobile. Only forward engineering and management commits were identified in the data, suggesting that high quality code was contributed and no major re-factoring was required for this plug-in. This result also supports our previous hypothesis that modular plug-in based OSS communities provide an efficient way for proprietary companies to participate and contribute with new functionality as new plug-ins. 

\textbf{Conclusion:} Decoupling of plug-ins helps in targeting contributions and quality improvement suggestions and simplifies the collaboration networks for discussions on bugs and future improvements.  



\section{Qualitative Analysis}
\label{sec:Qualitativeanalysis}
We conducted thematic analysis \cite{Cruzes14,Cruzes2011440} to find recurring patterns in the collected qualitative data. The following steps were performed in the process.
\begin{enumerate}
\item Transcribe the interviewed data from the five interviewee (see Table \ref{tab:interviewees}).
\item Identify and define five distinct themes in the data  (see Table \ref{tab:themes}).
\item Classify the interview statements based on the defined themes.
\item Summarize the findings and answers to the RQs.
\end{enumerate}

\begin{table*}[t]
  \label{tbl:Themes}
  \centering
  \caption{Themes emerging from the thematic analysis.}
    \begin{tabular}{p{5cm} p{6cm}}
    \textbf{Theme name} &
      \textbf{Definition}
      \\
       \toprule
    \textbf{Opening up} &
      Sony Mobile's transition process from closed innovation model to OI model.
      \\
      \midrule
    \textbf{Determinants of openness} &
      Factors that Sony Mobile considers before indulging themselves into OI.
      \\
      \midrule
    \textbf{Requirements engineering} &
      How Sony Mobile manages their requirements while working in OI context.
      \\
      \midrule
    \textbf{Testing} &
      How Sony Mobile manages their testing process while working in OI context.
      \\
      \midrule
    \textbf{Innovation outcome} &
      The outcomes for Sony Mobile as a consequence of adopting OI.
      \\
      \bottomrule
    
    \end{tabular}%
  \label{tab:themes}%
\end{table*}%
\subsection{Opening up}
\label{sec:Triggers}

The process of opening up for external collaboration and maturing as an open source organization, can be compared to moving from a closed innovation model to an OI model~\cite{chesbrough2006}. The data suggest that the trigger for this process was a paradigm shift around 2010 when Sony Mobile moved from the Symbian platform (developed in a joint venture), to Google's open source Android platform in their products~\cite{west2013evolving}. Switching to Android correlates to a general shift in the development environment, moving from Windows to Linux. This concerned the tools used in the product development as well. A transition was made from existing proprietary solutions, e.g. the build-server Electric commander, to the tools used by Google in their Android development, e.g. GIT and Gerrit. As stated by I2, \textit{``\ldots suddenly we were almost running pretty much everything, at least anything that touches our phone development, we were running on Linux and open source''}. This was not a conscious decision from management but rather something that grew bottom-up from the engineers. The engineers further felt the need for easing off the old and complex chain of integration and building process. 

At the same time, a conscious decision was made regarding to what extent Sony Mobile should invest in the open source tool chain. As  stated by I5, \textit{``\ldots not only should [the tool chain] be based on OSS, but we should behave like an active committer in the ways we can control, understand and even steer it up to the way we want to have it''}. The biggest hurdle concerned the notion of giving away internally developed intellectual property rights, which could represent competitive advantage. The legal department needed some time to understanding the benefits and license aspects, which caused the initial contribution process to be extra troublesome. In this case, Sony Mobile benefited from having an internal champion and OSS evangelist (I5). He helped to drive the initiative from the management side, explained the issues and clarified concerns from different functions and levels inside Sony Mobile. Another success factor was the creation of an OSS review board, which included different stakeholders such as legal department representatives, User Experience (UX) design, product development and product owners. This allowed for management, legal, and technology representatives to meet and discuss OSS related issues. The OSS contribution process now includes submitting a form for review, which promotes it further after successful initial screening. Next, the OSS review board gives it a go or no-go decision. As this would prove bureaucratic if it would be needed for each and every contribution to an OSS community, frame-agreements are created for open source projects with a high-intensity involvement, e.g. Jenkins and Gerrit. This creates a simplified and more sustainable process allowing for a day to day interaction between developers in the Tools department and the communities surrounding Jenkins and Gerrit. Sony Mobile's involvement in OSS communities is in-line with the findings of governance in OSS communities by Jensen~\cite{jensen2010governance}.

\textbf{Conclusion:} Adopting OI was a result of a paradigm shift moving from Windows to Linux environment to stay as close as possible to Google's tool chain. Furthermore, Sony Mobile saw a great potential in contributing to OSS communities (Jenkins and Gerrit) and steering them towards its own organizational interests, as opposed to buying costly proprietary tools.

\subsection{Determinants of Openness}
\label{sec:openness}
Several factors interplay in the decision process of whether or not a feature or a new project should be made open.
Jenkins and Gerrit are neither seen as a part of Sony Mobile's competitive advantage nor as a source of revenue. This is the main reason why developers in the Tools department can meet with competitors, go to conferences, give away free work etc. This, in turn, builds a general attitude that when something is about to be created, the question asked beforehand is if it can be made open source. There is also a follow-up question, whether Sony Mobile would benefit anything from it, for example maintenance, support and development from an active community. If a feature or a project is too specific and it is deemed that it will not gain any traction, the cost of generalizing the project for open use is not motivated. Another factor is whether there is an existing community for a feature or a project. By contributing a plug-in to the Jenkins community or a feature to Gerrit there is a chance that an active workforce is ready to adopt the contribution, whilst for new projects this has to be created from scratch which may be cumbersome.

Another strategic factor concerns having a first-mover advantage. Contributing a new feature or a project first means that Sony Mobile as the maintainer gets a higher influence and a greater possibility to steer it in their own strategic interest. If a competitor or the community publishes the project, Sony Mobile may have less influence and will have to adapt to the governance and requirements from the others. A good example here is the Gerrit-trigger. 
The functionality was requested internally at Sony Mobile and therefore undergone development by the Tools department during the same period it became known that there was a similar development ongoing in the community. As stated by I3, \textit{``\ldots we saw a big risk of the community going one way and us going a very different route''}. This led to the release of the internal Gerrit-trigger as an open source plug-in to Jenkins, which ended up being the version with gained acceptance in the Jenkins and Gerrit communities. The initial thought was however to keep it closed according to I3, \textit{``\ldots We saw the Gerrit-trigger plug-in as a differentiating feature meaning that it was something that we shouldn't contribute because it gave us a competitive edge towards our competitors [in regards to our continuous integration process]''}. It should be noted that this was in the beginning of the process of opening up in Sony Mobile and a positive attitude was rising.
A quote from I3 explains the positive attitude of the organization which might hint about future directions, \textit{``\ldots in 5 years’ time probably everything that Sony Mobile does would become open''}.

\textbf{Conclusion:} One of the key determinants of making a project open is that it is not seen as a main source of revenue. In other words, there is no competitive advantage gained by Sony Mobile by retaining the project in-house. By maintaining an internal fork, the project incurs more maintenance cost compared to making it open source. Therefore, all the all projects with no competitive advantage are seen as good candidates to become open source.

\subsection{Requirements Engineering}
\label{sec:requirementsandOpeninnovation}

This theme provides insights about requirements engineering practices in an example OI context. The requirements process in the Tools department towards the Jenkins and Gerrit communities does not seem very rigid, which is a common characteristic for OSS~\cite{scacchi2002}. The product development teams in Sony Mobile are the main customers of the Tools department. The teams are, however, quite silent with the exception of one or two power users. 
There is an open backlog for internal use inside Sony Mobile where anyone from the product development may post feature requests. However, a majority of the feature requests are submitted via e-mail. The developers in the Tools department started arranging monthly workshops where they invited the power users and the personnel from different functional roles in the product development organization. An open discussion is encouraged allowing for people to express their wishes and issues. An example of an idea sprung out from this forum is the Build-failure-analyzer\footnote{https://wiki.jenkins-ci.org/display/JENKINS/BuildFailureAnalyzer} plug-in. Most of the requirements are, however, elicited internally within the Tools department in a dialogue between managers, architects and developers. They are seen to have the subject matter expertise in regards to the tool functionality. According to I2, there are \textit{``\ldots architect groups which investigate and collaborate with managers about how we could take the tool environment further''}. This is formulated as focus areas, and \textit{``\ldots typical examples of these requirements are sync times, push times, build times and apart from that everything needs to be faster and faster''}. These requirements are high level and later delegated to the development team for refinement. 

The Tools team works in an agile Scrum-like manner with influences from Kanban for simpler planning. The planning board contains a speed lane which is dedicated for severe issues that need immediate attention. The importance of being agile is highlighted by I2, \textit{``\ldots We need to be agile because issues can come from anywhere and we need to be able to react''}. 

The internal prioritization is managed by the development team itself, on delegation from the upper manager, and lead by two developers which have the assigned role of tool managers for Jenkins and Gerrit respectively. The focus areas frame the areas which need extra attention. Every new feature is prioritized against existing issues and feature requests in the backlog. External feature requests to OSS projects managed by the Tools department (e.g. the Gerrit-trigger plug-in) are viewed in a similar manner as when deciding whether to make an internal feature or project open or not. If it is deemed to benefit Sony Mobile enough, it will be put in the backlog and it will be prioritized in regards to everything else. As stated by I3, \textit{``\ldots We almost never implemented any feature requests from outside unless we think that it is a good idea [for Sony Mobile]''}. 
If it is not interesting enough but still a good idea, they are open for commits from the community.

An example regards the Gerrit-trigger plug-in and the implementation of different trigger styles. Pressing issues in the Tools department's backlog kept them from working on the new features. At the same time, another software intense organization with interest in the plug-in contacted the Tools department about features they wanted to implement. These features and the trigger style functionality required a larger architectural reconstruction. It was agreed that the external organization would perform the architectural changes with a continuous discussion with the Tools department. This allowed for a smaller workload and the possibility to implement this feature earlier. This feature-by-feature collaboration is a commonly occurring practice as highlighted by I1, \textit{``It's mostly feature per feature. It could be an organization that wants this feature and then they work on it and we work on it". But we don't have any long standing collaborations''}. I3 elaborates on this further and states that \textit{``\ldots it is quite common for these types of collaboration to happen just between plug-in maintainer and someone else. They emailed us and we emailed back''} as was the case in the previous example.

In the projects where the Tools department is not a maintainer, community governance needs more care. In the Gerrit community, new features are usually discussed via mailing lists. However, large features are managed at hackathons by the Tools department where they can communicate directly with the community to avoid getting stuck in tiny details~\cite{Morgan2012}. As brought up by I2, \textit{``\ldots with the community you need to get people to look at it the same way as you do and get an agreement, otherwise it will be just discussions forever''}. This is extra problematic in the Gerrit community as the inner core team with the merge rights consists of only six people, of which one is from Sony Mobile. One of the key features received from the community was the tagging support for patch sets. I2 stated, \textit{``\ldots When developers upload a change which can have several revisions, it enabled us to tag meta-data like what is the issue in our issues handling system and changes 
in priorities as a result of that change. This tagging feature allows the developers to handle their work flow in a better way"}. This whole feature was proposed and integrated during a hackathon, and contained more than 40 shared patch sets. Prior to implementing this feature together with the community (I3 quoted) \textit{``\ldots we tried to do it with the help of external consultants but we could not get it in, but meeting core developer in the community did the job for us"}. 

As hackathons may not always be available, an alternative way to communicate feature suggestions more efficiently is by mock-ups and prototypes. I3 described how important it is to sell your features and get people excited about it. Screenshots is one way to visualize it and show how it can help end-users. In the Jenkins community, this has been taken further by hosting official webcasts where everyone is invited to present and show new development ideas. Apart from using mailing lists and existing communication channels, Sony Mobile creates their own channels, e.g. with public blogs aimed at developers and the open source communities. 

This close collaboration with the community is important as Sony Mobile does not want to end up with an internal fork of any tool. An I2 quoted, \textit{``If we start diverging from the original software we can't really put an issue in their issue tracker because we can't know for sure if it's our fault or their system and we would loose the whole way of getting help from community to fix stuff and collaborate on issues''}. Another risk would be that \textit{``\ldots all of a sudden everybody is dependent on stuff that is taken away from the major version of Gerrit. We cannot afford to re-work everything''}. Due to these reasons, the Tools department is keen on not keeping stuff for themselves, but contributing everything~\cite{ven2008challenges, Wnuk2012}.
An issue in Jenkins is that there exist numerous combinations and settings of plug-ins. Therefore, it is very important to have backward compatibility when updating a plug-in and planning new features.

\textbf{Conclusion:} The requirements engineering process does not seem to be very rigid, and a majority of the features requests are submitted through e-mails, and monthly workshops with the power users (e.g. internal developers and testers). However, large features are discussed directly with the community at hackathons by the Sony Mobile's Tools department to avoid communication bottlenecks. Furthermore, the prioritization of features is based on the internal needs of Sony Mobile.

\subsection{Testing}
\label{sec:testingandOpeninnovation}

Similar to the requirements process, the testing process does not seem very rigid either. I3 quoted, \textit{``\ldots When we fix something we try to write tests for that so we know it doesn't happen again in another way. But that's mostly our testing process I think. I mean, we write JUnit and Hudson test cases for bugs that we fix''}. 

Bugs and issues are, similarly to feature requests, reported internally either via e-mail or an open backlog. Externally, bugs or issues are reported via the issue trackers available in the community platforms. The content of the issue trackers is based on the most current pressing needs in the Tools department. Critical issues are prioritized via the Kanban speed lane which refers to a prioritized list of requirements/bugs based on the urgent needs of Sony Mobile. If a bug or an issue has low priority, it is reported to the community. This self-focused view correlates with the mentality of how the organization would benefit from making a certain contribution, which is described to apply externally as well, \textit{``\ldots Organizations take the issues that affect them the most''}. However, it is important to show to the community that the organization wants to contribute to the project as a whole and not just to its parts, as mentioned by Dahlander~\cite{Dahlander20061243}. In order to do so, the Tools department continuously stays updated about the current bugs and their status. It is a collaborative work with giving and taking. \textit{``Sometimes, if we have a big issue, someone else may have it too and we can focus on fixing other bugs so we try to forward as many issues as possible''}.
 
In Gerrit, the Tools department is struggling with an old manual testing framework. Openness has lead them to think about switching from the manual to an automated testing process. I2 stated, \textit{``\ldots It is one of my personal goals this year to figure out how we can structure our Gerrit testing in collaboration with the community. Acceptance tests are introduced greatly 
in Gerrit too but we need to look into and see how we can integrate our tests with the community so that the whole testing becomes automated''}. In Jenkins, one of the biggest challenges in regards to test is to have a complete coverage as there are many different configurations and setups available due to the open plug-in architecture. However, Gerrit still has some to catch up as stated by I2, \textit{``it is complex to write stable acceptance tests in Gerrit as we are not mature enough compared to Jenkins''}. A further issue is that the test suites are getting bigger and therefore urges the need for automated testing. 

Jenkins is considered more mature since the community has an automated test suite which is run every week when a new version of the core is released. This test automation uses  Selenium\footnote{http://www.seleniumhq.org/}, which is an external OSS test framework used to facilitate the automated acceptance tests. It did not get any traction until recently because it was written in Ruby, while the Jenkins community is mainly Java-oriented. This came up after a discussion at a hackathon where the core members in the community gathered, including representatives from the Tools department. It was decided to rework the framework to a Java-based version, which has helped the testing to take off although there still remains a lot to be done. 

I3 highlighted that Sony Mobile played an important role in the Selenium Java transition process, \textit{``The idea of an acceptance test harness came from the community but [Sony Mobile] was the biggest committer to actually getting traction on it''}. From Sony Mobile's perspective, it can contribute its internal acceptance tests to the community and have the community execute what Sony Mobile tests when setting up the next stable version. Consequently, it requires less work of Sony Mobile when it is time to test a new stable version. From the community perspective I3 stated, \textit{``an Acceptance Test Harness also helps the community and other Organizations to understand what problems that big or small organizations have in terms of features or in terms other requirements on the system. So it's a tool where everyone helps each other''}.

\textbf{Conclusion:} Like the requirements engineering process, the testing process is also very informal, and Sony Mobile prioritizes the issues that affect them the most. One of the biggest challenges faced by the community and organizations is to have complete test coverage due to the open plug-in architecture. The introduction of an acceptance test harness was an important step to make the whole testing process automated for organizations, and the Jenkins and Gerrit communities.

\subsection{Innovation Outcomes}
\label{sec:innovationoutcome}
The word \textit{innovation} has a connotation of newness~\cite{assink2006inhibitors} and can be classified as either things (products and services), or changes in the way we create and deliver products, services and processes. Assink~\cite{assink2006inhibitors} classified innovation into disruptive and incremental. Disruptive innovations change the game by attacking an existing business and offering great opportunities for new profits and growth. Incremental innovations remain within the boundaries of the existing technology, market and technology of an organization. The innovation outcomes found in this study are related to incremental innovations. 

Sony Mobile does not have any metrics for measuring process and product innovation outcomes. However, valuable insights were found during the interviews regarding what Sony Mobile has gained from the Jenkins and Gerrit community involvement. During the analysis, the following innovation outcomes have been identified:
\begin{enumerate}
 \item Free features.
 \item Free maintenance.
 \item Freed-up time.
 \item Knowledge retention.
 \item Flexibility in implementing new features and fixing bugs.
 \item Increased turnaround speed.
 \item Increased quality assurance.
 \item Improved new product releases and upgrades.
 \item Inner source initiative.
 \end{enumerate} 
The most distinct innovation outcome is the notion of obtaining \textbf{\textit{free features}} from the community, which have different facets~\cite{Dahlander2008, stuermer_extending_2009}. For projects maintained by Sony Mobile, such as the Gerrit-trigger plug-in, a noticeable amount of external commits can be accounted for. Similarly, in communities where Sony Mobile is not a maintainer, they can still account for free work, but it requires a higher effort in lobbying and actively steering the community in order to maximize the benefits for the organization. Along also comes, the \textbf{\textit{free maintenance}} and quality assurance work, which renders better quality in the tools. Furthermore, the use of tools (Jenkins and Gerrit) helped software developers and testers to better manage their work-flow. Consequently, it \textbf{\textit{freed-up time}} for the developers and testers that could be used to spent on other innovation activities. The observed innovation example in this case was the developers working with OSS communities, acquiring and integrating the external knowledge into internal product development. 

Correlated to the \textbf{\textit{free work}} is the acknowledgement that the development team of six people in the Tools department will have a hard time keeping up with the external workforce, if they were to work in a closed environment. \textit{``\ldots I mean Gerrit has like let us say we have 50 active developers, it's hard for the tech organization to compete with that kind of workforce and these developers at Gerrit are really smart guys. It is hard to compete for commercial Organizations''}. Further on, \textit{``\ldots We are mature enough to know that we lose the competitive edge if we do not open up because we cannot keep up with hundreds of developers in the community that develops the same thing''}.

An organizational innovation outcome of opening up is the \textbf{\textit{knowledge retention}} which comes from having a movable workforce. People in the community may move around geographically, socially and professionally but can still be part of the community and continue to contribute. I3, who took part in the initiation of many projects, recently left Sony Mobile but is still involved in development and reviewing code for his former colleagues which is in line with the findings of previous studies~\cite{Morgan2012,stuermer_extending_2009}. Otherwise, the knowledge tied to I3 would have risked being lost for Sony Mobile. 


Sony Mobile had many proprietary tools before opening up. Adapting these tools, such as the build server Electric commander, was cumbersome and it took long time before even a small fix would be implemented and delivered by the supplier. This created a stiffness whereas open source brought \textbf{\textit{flexibility}}. I2 quoted, \textit{``\ldots Say you just want a small fix, and you can fix that yourself very easily but putting a requirement on another organization, I mean it can take years. Nothing says that they have to do it''}. This increase in the \textit{\textbf{turnaround speed}} was besides the absence of license fees, a main argument in the discussions when looking at Jenkins as an alternative to Electric commander. This was despite the required extra involvement and cost of more internal man-hours. As a result, the continuous integration tool chain could be tailored specifically to the needs of the product development team. I1 stated that \textit{``\ldots Jenkins and Gerrit have been set up for testers and developers in a way that they can have their own projects that build code and make changes. Developers can handle all those parts by themselves and get to know in less than 3 minutes whether or not their change had introduced any bugs or errors to the system"}. Ultimately, it provides \textit{\textbf{quality assurance}} and performance gains by making the work flow easier for software developers and testers. Prior to the introduction of these tools there was one engineer who was managing the builds for all developers. In the current practice everybody is free to extend on what is given to them from tools department. It offers more scalability and flexibility~\cite{Morgan2010}. 

I1 stated that besides the flexibility, the Tools department is currently able to make a \textit{``\ldots more stable tools environment [at Sony Mobile] 
and that sort of makes our customers of the tools department, the testers and the engineers, to have an environment that actually works and does not collapse while trying to use it''}. I2 mentioned that \textit{``\ldots I think it is due to the part of open source and we are trying to embrace all these changes to our advantage. I think we can make high quality products in less time and in the end it lets us make better products. I think we never made an as good product as we are doing today''}. Further exploration of this statement revealed the background context where Sony Mobile has \textit{\textbf{improved}} in terms of handling all the \textbf{\textit{new releases and upgrades}} in their phones compared to their competitors and part of its credit is given to the flexibility offered by the open source tools Jenkins and Gerrit. 

The obtained external knowledge about the different parts of the continuous integration tool chain enabled better product development. However, the Tools department has to take the responsibility for the whole tool chain and not just its different parts, e.g. Jenkins and Gerrit, described by I5 as the next step in the maturity process. The tool chain has the potential to function as an enabler in other contexts as well, seeing Sony Mobile as a diversified organization with multiple product branches. By opening up in the way that the Tools department has done, effects from the coupled OI processes with Jenkins and Gerrit may spread even further into other product branches, possibly rendering in further innovations on different abstraction levels~\cite{LinakerSurvey15}. A way of facilitating this spread is the creation of an \textit{\textbf{inner source initiative}} which will allow for knowledge sharing across the different borders inside Sony Mobile, comparable to an internal OSS community, or as a bazaar inside a cathedral~\cite{wesselius2008}. The tool chain is even seen as the foundation for a platform which is supposed to facilitate this sharing~\cite{linaaker2014}. The Tools department is considered more mature in terms of contributing and controlling the OSS communities. Hence, the Tools department can be used as an example of how other parts of the organization could open up and work with OSS communities. I5 uses this when evangelizing and working on further opening up the organization at large, and describes how \textit{``\dots they've been spearheading the culture of being active or in engaging something with communities''}.

\textbf{Conclusion:} Some of the innovation outcomes attached to Sony Mobile's openness entail more freed-up time for developers, better quality assurance, improved product releases and upgrades, inner source initiatives and faster time to market.

\section{Mapping between sections and answers to research questions}
\label{sec:RQmapping}
Table~\ref{tab:RQmapping} presents the mapping of research questions to answers with section numbers. Furthermore, a brief summary of answers to research questions is highlighted in section~\ref{sec:conclusion}. 

\begin{table}[htbp]
  \centering
  \caption{Mapping of answers to RQs with section numbers}
    \begin{tabular}{ll}
    \textbf{Research questions} & \multicolumn{1}{l}{\textbf{Answers to RQs}} \\ \toprule
    RQ1   & Section~\ref{sec:RQ1InvolvementofSony}  \\ \midrule
    RQ2   &  Section~\ref{sec:RQ2Openingup},~\ref{sec:RQ2OpennessvsProprietarysoftware} \\ \midrule
    RQ3   &  Section~\ref{sec:RQ3DeterminantsofOpenness}\\ \midrule
    RQ4   &  Section~\ref{sec:RQ4Innovation Outcomes}\\ \midrule
    RQ5   &  Section~\ref{sec:RQ5RequirementsEngineering},~\ref{sec:RQ5Testing}\\ \bottomrule
    \end{tabular}%
  \label{tab:RQmapping}%
\end{table}%

\section{Results and Discussion}
\label{sec:discussion}
Results from the quantitative and qualitative analysis are discussed below, of which the latter is addressed per theme, and connected to the research questions defined in Table \ref{tab:Researchquestions}.

\subsection{Involvement of Sony Mobile in OSS Communities}
\label{sec:RQ1InvolvementofSony}
Addressing \textbf{RQ1} in Table~\ref{tab:Researchquestions}, the quantitative analysis showed that Sony Mobile has an active role in numerous OSS projects. In most of the analysed projects, Sony Mobile is the initiator and maintainer. An exception is Gerrit where they entered an already established project. However, with 8.2 \% (see Table~\ref{tab:Sonymobilecontributionpercentages}) of the commits during the investigated time-span, they have established themselves in the community and been able to contribute the necessary adaptions for Gerrit to function as a part of the continuous integration tool-chain used inside Sony Mobile. This shows that Sony Mobile has an open mindset to creating their own OSS projects, as well as getting involved and contributing back in existing ones. In the projects which Sony Mobile has released themselves, they further show that they are open for contributions by others. In the Gerrit-trigger plug-in for example, they only represent 65\% of the total commits. This also gives a clear picture of the help gained by the external workforce as highlighted by OI. By opening up the Gerrit-trigger plugin and making it a part of the Jenkins community, they earn benefits such as shared feature development, maintenance and quality assurance. A reason why some of the other projects have fewer external commits (e.g., PyGerrit, Build-failure-analyzer and Team-views) may be that they are not as established and attractive for others outside Sony Mobile. A further explanation could be that Sony Mobile has not invested the time and attention needed in order to build successful communities around these projects. 


\subsection{Opening Up}
\label{sec:RQ2Openingup}
In relation to \textbf{RQ2}, the move to Android took Sony Mobile from a closed context to an external arena for OI, recalls the description provided by Grotnes~\cite{grotnes2009standardization}. With this, the R\&D was moved from a structured joint venture and an internal vertical hierarchy to an OI community. This novel way of using pooled R\&D~\cite{ west2006challenges} can be further found on the operational level of the Tools department, which freely cooperates with both known and unknown partners in the Jenkins and Gerrit communities. From the OI perspective, these activities can be seen as a number of outside-in and inside-out transactions. 

The Tools department's involvement in Jenkins and Gerrit and the associated contribution process are repetitive and bidirectional. Thus, this interaction can be classified as a coupled innovation process~\cite{gassmann_towards_2004}. This also complies with Grotnes' description of how an open membership renders in a coupled process, as Jenkins and Gerrit communities both are free for anyone to join, in contrast to the Android platform and its Open Handset Alliance, which is invite-only~\cite{grotnes2009standardization}.

The quantitative results provide further support for the hypothesis that both established, larger corporations  
and small scale software organizations are involved in the development of Jenkins and Gerrit (see Table \ref{tab:Sonymobilecontributionpercentages}). Some of the small organizations are Garmin,	Ostrovsky, Luksza,	Codeaurora,	Quelltextlich etc.
 This confirms findings from the existing OI literature, e.g. ~\cite{Stam2009,henkelchampions2008} that other community players also can use these communities as external R\&D resources and complimentary assets to internal R\&D processes. One possible motivation for start-ups or small scale organizations to utilize external R\&D is their lack of in-house R\&D capabilities. Large scale software organizations exploit communities to influence not only the development direction, but also to gain a good reputation in the community as underlined by prior studies~\cite{Dahlander20061243,henkelchampions2008}. 

Gaining a good reputation requires more than just being an active committer. Stam~\cite{Stam2009} separates between technical (e.g. commits) and social activities (e.g. organizing conferences and actively promoting the community), where the latter is needed as complementary in order to maximize the benefits gained from the former. Sony Mobile and the Tools department have evolved in this vein as they are continuously present at conferences, hackathons and in online discussions. Focused on technical activities, the Tools department have progressively moved from making small to more substantial commits. Along with the growth of commits, they have also matured in their commit strategy. As described in Section \ref{sec:openness}, the intent was originally to keep the Gerrit-trigger plug-in enclosed. This form of selective revealing~\cite{henkel_selective_2006} has however been minimized due to a more open mindset. As a consequence of the openness more plug-ins were initiated and the development time was reduced. 

Although the adoption of Jenkins and Gerrit came along with an adaption to the Android development, it was also driven bottom-up by the engineers since they felt the need for easing off the complex integration tool chain and building process as mentioned by Wnuk et al.~\cite{Wnuk2012}. As described in Section \ref{sec:Triggers}, this process was not free of hurdles, one being the cultural and managerial aspect of giving away internally developed intellectual property~\cite{Husig2011}. The fear to reveal intellectual property was resolved thanks to the introduction of an OSS review board that involved both legal and technical aspects. Having an internal champion to give leverage to the needed organizational and process changes, convince skeptical managers~\cite{henkelchampions2008}, and evangelism of open source was a great success factor, also identified in the inner source literature~\cite{lindman2008applying}. 


\subsection{Determinants of Openness}
\label{sec:RQ3DeterminantsofOpenness}
When discussing if something should be made open or closed (\textbf{RQ3}) in Table~\ref{tab:Researchquestions}, an initial distinction within the Tools department regarding the possible four cases is made: 

\begin{enumerate}
\item New projects created internally (e.g. Gerrit-trigger).
\item New features to non-maintained projects (e.g. Gerrit).
\item External feature requirement requests to maintained projects (e.g. Gerrit-trigger).
\item External bug reports to already maintained projects (e.g. Gerrit-trigger).
\end{enumerate} 

The first two may be seen as an inside-out transaction, whilst the two latter are of an outside-in character. All have their distinct considerations, but one they have in common, as described in Section \ref{sec:openness}, is whether Sony Mobile will benefit from it or not. Even though the transaction cost is relative low, it still needs to be prioritized against the current needs. In the case of the two former, if a feature is too specific for Sony Mobile's case it will not gain any traction, and it will be a lost opportunity cost~\cite{lerner2002some}.

The fact that Sony Mobile considers their supportive tools, e.g. Jenkins and Gerrit, as a non-competitive advantage is interesting as they constitute an essential part of their continuous integration process, and hence the development process. As stated in regards to the initial intent to keep Gerrit-trigger internally, they saw a greater benefit in releasing it to the OSS community and having others adopt it than keeping it closed. The fear that the community was moving in another direction, rendering in a costly need of patch-sets and possible risk of an internal fork, was one reason for giving the plug-in to the community~\cite{ven2008challenges}. Wnuk et al.~\cite{Wnuk2012} reason in a similar manner in their study where they differentiate between contributing early or late to the community in regards to specific features. By going with the former strategy, one may risk losing the competitive edge, however the latter creates potentially high maintenance costs.  

Sony Mobile is aware that increased mobility~\cite{chesbrough2006} poses a threat to the Tools department as it is not possible for them to work in the OSS communities' pace due to the limited amount of resources~\cite{chesbrough2006}. Consequently, it may end up damaging the originally perceived competitive advantage by lagging behind. On the other hand, openness gives Sony Mobile an opportunity to have an access to pragmatic software development workforce and also, Sony Mobile does not have to compete against the community. Additionally, by adopting a first mover strategy~\cite{lieberman1998first} Sony Mobile can use their contributions to steer and influence the direction of the community.


\subsection{Requirements Engineering}
\label{sec:RQ5RequirementsEngineering}
Tracing back to \textbf{RQ5} in Table~\ref{tab:Researchquestions}, the Tools department may be viewed as both a developer and an end-user, making up a source of requirements as can often be seen in Open Source Software Development (OSSD)~\cite{scacchi2002}. This applies both internally (as a supplier and an administrator of the tools), and externally (as a member of the communities). From an RE perspective, they are their own stakeholder, competing with other stakeholders (members) in the Jenkins and Gerrit communities.
These are important characteristics as stakeholders who are not developers are often neither identified nor considered~\cite{alspaugh2013ongoing}. A consequence otherwise could be that certain areas are forgotten or neglected which stands in contrast to Wnuk et al.~\cite{Wnuk2012} who state that adoption of OI makes identifying stakeholders' needs more manageable. Further, this brings an interesting contrast to traditional RE where non-technical stakeholders often need considerable help in expressing themselves. The RE in OI applied through OSS can be seen as quicker, light-weight and more technically oriented than traditional RE~\cite{scacchi2002}. 



In OSSD, one often needs to have a high authority level or have a group of stakeholders backing up the intent. Sony Mobile has been very successful in this respect due to the Tools department involvement inside these communities~\cite{Dahlander20061243}. Due to their high commitment and good track record, Sony Mobile employees have reached a high level  
in the governance organization. The Tools department combines these positions in the communities together with openness in terms of helping competitors and interacting in social activities~\cite{Stam2009} (e.g. developer conferences~\cite{knauss2014openness}). One reason for this is to attract quiet stakeholders, both in terms of influencing the community~\cite{Dahlander2008}, but also to get access to others' knowledge which could be relevant for Sony Mobile. An example of this is the introduced focus on scalability in both the Jenkins and Gerrit communities, where the Tools department needed to find stakeholders with similar issues to raise awareness and create traction to the topic. Communication in this requirements value chain~\cite{fricker2010requirements} between the different stakeholders, as well as with grouping can be deemed very ad-hoc, similar to OSS RE in general~\cite{scacchi2002}. This correlates to the power structure and how influence may move between different stakeholders.

Social interaction between the stakeholders is stressed by Panjer et al.~\cite{panjer2008cooperation} as an important aspect to resolve conflicts and to coordinate dependencies in distributed software development projects. The Tools department's preference for live meetings over the otherwise available electronic options such as mailing lists, issue trackers and discussion boards, is due to time differences and lag in discussions that complicate implementation of larger features. Open source hackathons~\cite{scacchi2010collaboration} is the preferable choice as it brings the core stakeholders together which allows for informal negotiations~\cite{fricker2010requirements} and a live just-in-time requirements process~\cite{ernst2012case}, meaning that requirements are captured in a less formal matter and first fully elaborated during implementation. As highlighted in Section~\ref{sec:requirementsandOpeninnovation}, feature-by-feature collaborations is also a common practice. This is also due to the ease of communication as it may be performed between two single parties. Hence, it may be concluded that communication in this type of distributed development is a critical challenge, and in this case overcome by live meetings and keeping the number of collaborators per feature low.

This use of live-meetings and social events for requirements communication and discussion, highlights the importance of being socially present in a community other than just online if a stakeholder wants to stay aware of important decisions and implementations. Another reason for the individual stakeholder is to maintain or grow its influence and position in the governance ladder. Hence, organizations might need to revise their community involvement strategy and value what their intents are in contrast to if an online presence is enough.

Another interesting reflection on the feature-by-feature collaborations is that these may be performed with different stakeholders, i.e. relations between stakeholders fluctuate depending on their respective interests. This objective and short-term way of looking at collaborations imply a need of standardized practices in a community for it to be effective.

\subsection{Testing}
\label{sec:RQ5Testing}
Addressing the \textbf{RQ5} in Table~\ref{tab:Researchquestions}, we noticed during interviews that both Jenkins and Gerrit focus on manual test cases. At the same time, the communities started the transformation journey towards automated testing, with the Jenkins community leading. The openness of the Tools department led them to participate in the testing part of Jenkins community and to use its influence to rally the traction towards it amongst the other stakeholders in the community. This is especially important for the Jenkins community due to the rich number of settings offered by the plug-ins.

The Gerrit community is currently following the Jenkins' community patch, as stressed by interviewee I2. With this move towards automated testing, quality assurance will hopefully become better and enable more stable releases. These are important aspects and business drivers for the Tools department as Jenkins and Gerrit constitute the critical parts in Sony Mobile's continuous integration tool chain. From this perspective, a trend may be seen in how the different communities are becoming more professionalized in the sense that the tools make up business critical assets for many of the stakeholders in the communities, which motivates a continuous effort in risk-reduction~\cite{RedHatIBM2009,henkel_selective_2006}.

The move towards automated testing also allowed for the Tools department to contribute their internal test cases. This may be viewed as profitable from two angles. First, it reduces the internal workload and second, it secures that settings and cases specific for Sony Mobile are addressed and cared for. The test cases may to some extent be viewed as a set of informal requirements, which secure quality aspects in regards to scalability for example which is important for Sony Mobile~\cite{Bjarnason2015}. Similar practices, but much more formal, are commonly used in more traditional (closed) software development environments. From a community perspective, other stakeholders benefit from this as they get the view and settings from a large environment which enable them to grow as well.

As can be noted in Table~\ref{fig:bubbleplot}, the focus is on forward and re-engineering. An interesting concern is when and how much one should contribute to bug fixes and what should be left for the community, because some bug fixes are very specific to Sony Mobile and the community will not gain anything from them. As discussed earlier, Sony Mobile has the strategy of focusing on issues which are self-beneficiary. Therefore, to be able to keep the influence and strategic position in the communities, the work still has to be done in this area as well. 


\subsection{Innovation Outcomes}
\label{sec:RQ4Innovation Outcomes}
In relation to \textbf{RQ4} in Table~\ref{tab:Researchquestions}, the focal point of the OI theory is value creation and capture~\cite{chesbrough_open_2003}. In the studied case, the value is created and captured through their involvement in the Jenkins and Gerrit communities. However, measuring that value using key performance indicators is a daunting challenge. Edison et al.~\cite{Edison2013} confirmed a limited number of measurement models, and that the existing ones neither model all innovation aspects, nor say what metric can be used to measure a certain aspect. Furthermore, existing literature is scarce in regards to how data should be gathered and used for the metrics proposed in the literature. As expected, we found that Sony Mobile does not have established mechanisms in place to measure their performance before and after the Jenkins and Gerrit introduction. 
However, from the qualitative data collected from the interviews we specifically looked for two types of innovations: product innovations in the tools Jenkins and Gerrit, and process innovation in Sony Mobile's product development. Other types, specifically market and organizational innovation were considered but not identified.

By taking an active part in the knowledge sharing and exchange process with communities~\cite{Dahlander2008, stuermer_extending_2009}, the Tools department enjoys the benefits of contributions extending the functionality of their continuous integration tools. Another benefit is the free maintenance and bug corrections and the test cases extension for further quality assurance. By extension, these software improvements may be labeled as product innovations depending on what definition to be used~\cite{Edison2013}. This may also be viewed from the process innovation perspective~\cite{osloManual2005} as Sony Mobile gets access to extra work force and a broad variety of competencies, which are internally unavailable~\cite{Dahlander2008}. The interviewees admit to that even a large scale software organization cannot keep up the technical work force beyond the organization's borders and there is a huge risk of losing the competitive edge by not being open. This is an acknowledgement to Joy's law~\cite{lakhani_principles_2007} \textit{``No matter who you are, not all smart people work for you''}. Hence, it is vital to reach work force beyond organisational boundaries when innovating~\cite{chesbrough_open_2003}, and knowledge is still retained even if people move around inside the community.

Furthermore, these software improvements and product innovations affect the performance and quality of the continuous integration process used by Sony Mobile's product development. Continuous integration as an agile practice~\cite{beck2001agile} enables early identification of integration issues as well as increases the developers' productivity and release frequency ~\cite{staahl2014modeling}. With this reasoning, as reported elsewhere~\cite{LinakerSurvey15}, we deem that the product innovations captured in Jenkins and Gerrit transfer on as process innovation to Sony Mobile's product development. The main reason behind this connection is the possibility to tailor and be flexible that OSS development permits. By adapting the tool chain to the specific needs of the product development the mentioned benefits (e.g. increased build quality and performance) are achieved and waste is reduced in the form of freed up hours, which product developers and testers may spend on alternative tasks, as confirmed by Moller~\cite{mollercritical2015}. Reduced time to market and increased quality of products are among the visible business outcomes. However, these outcomes cannot be confirmed due to a lack of objective metrics and came up as a result of interviews. 

Another process innovation, which could also be classified as an organizational innovation outcome~\cite{osloManual2005} is the inner source initiative. This initiative not only helps Sony Mobile to spread the tool chain, but also to build a platform (i.e. software forge~\cite{linaaker2014}) for sharing built on the tool within the other business units of Sony Mobile. This may be seen as an intra-organizational level OI as described by Morgan et al.~\cite{Morgan2012}. By integrating the knowledge from other domains, as well as opening up for development and commits, this allows a broader adoption and a higher innovation outcome for Sony Mobile and neighboring business units, as well as for communities. Organizational change in regards to processes and structures and related governance issues, would however be one of many challenges~\cite{Morgan2012}. Since Sony Mobile is a multinational corporation with a wide spread of internal culture, organizational changes are context and challenging.

\subsection{Openness of Tools Software vs. Proprietary software}
\label{sec:RQ2OpennessvsProprietarysoftware}
A specific aspect of \textbf{RQ2} in Table \ref{tab:Researchquestions} is that Sony Mobile only opens up its non-competitive tools that are not the part of the revenue stream. I3 stated that \textit{``\dots Sony Mobile has learnt that even collaborating with its worst competitors does not take away their competitive advantage, rather they bring help for Sony Mobile and becomes better and better''}. This raises a discussion point of why Sony Mobile limits its openness to noncompetitive tools, despite knowing that opening up creates a win-win situation for all stakeholders involved. Furthermore, it remains an open question why the research activity related to OI in SE is low, as confirmed by the results of a mapping study performed on the area~\cite{MunirMapping15}. 

In the light of the mapping study, it would be fair to state that the SE literature lacks studies on OI~\cite{MunirMapping15}. Organizations have a tendency to open proprietary products when they lose their value, and spinning off is a one way of re-capturing the value by creating a community around it~\cite{lindencommodification2009}. This implication paves the way for future studies using proprietary solutions as units of analysis. Moreover, it will lead to contextualization of OI practices, which may or may not work under different circumstances. Therefore, the findings could also be used to address the lack of contextualization weakness of OI mentioned by Mowery~\cite{mowery_plus_2009}. It is also important to note that this study focuses on OI via OSS participation, which is significantly different from the situation where OI is based on open source code for the product itself (like Android or Linux). In future work we plan to explore that situation to see if there are other patterns in these OI processes.

\section{Conclusions}
\label{sec:conclusion}
This study focuses on OI in SE at two levels: 1) innovation incorporated into Jenkins and Gerrit as software products, and 2) how these software improvements affect process and product innovation of Sony Mobile. By keeping the development of the tools open, the in- and out-flows of knowledge between the Tools department and the OSS communities bring improvement to Sony Mobile and innovate the way how products are developed. This type of openness should be separated from the cases where OSS is used as a basis for the organization's product or service offering, e.g. as a platform, component or full product~\cite{ven2008challenges}. To the best of our knowledge, no study has yet focused on the former version, which highlights the contribution of this study and the need for future research of the area.

Our findings suggest that both incumbents and many small scale organizations are involved in the development of Jenkins and Gerrit (\textbf{RQ1}). Sony Mobile may be considered as one of the top committers in the development of the two tools. The main trigger behind adopting OI turned out to be a paradigm shift, moving to an open source product platform \textbf{(RQ2)}. Sony Mobile's opening up process is limited to the tools that are non-competitive and non-pecuniary. Furthermore, Sony Mobile makes projects or features open source, which are neither seen as a main source of revenue nor as a competitive advantage \textbf{(RQ3)}.

In relation to the main innovation outcomes from OI participation (\textbf{RQ4}), we discovered that Sony Mobile lacks quantitative indicators to measure its innovative capacity before and after the introduction of OSS at the Tools department. However, the qualitative findings suggest that it has made the development environment more stable and flexible. One key reason, other than commits from communities, regards the possibility of tailoring the tools to internal needs. Still, it is left for future research efforts to further investigate in how OI adoption affects product quality and time to market. 

When looking at the impact of OI adoption on requirements and testing processes (\textbf{RQ5}), Sony Mobile uses dedicated internal resources to gain influence, which together with an openness toward direct competitors and communities is used to draw attention to issues relevant for Sony Mobile, e.g. scalability of tools to large production environments. Social presence outside of online channels is highly valued in order to manage communication challenges related to distributed development. Another way of tackling such challenges regards co-creation on a feature-by-feature basis between two single parties. Choice of partner fluctuates and depends on the feature in question and individual needs of the respective parties. Further, prioritization is made in regards to how an issue or feature may be seen as beneficial, in contrast to the pressing needs of the moment. Regarding testing, much focus is directed towards automating test activities in order to raise quality standards and professionalize communities to organizational standards.

The scope of the study findings is limited to software organizations with similar context, domain and size as Sony Mobile. It is also worth mentioning that the involvement of stakeholders in the Jenkins and Gerrit OSS communities suggests that the continuous integration processes of these OSS projects are comparable to the corresponding process at Sony Mobile. Thus, we believe that findings of this study may also be applicable to incumbents as well as small software organizations identified in this study. 

Future work includes investigation of other contexts and cases where companies use OSS aiming to leverage OI, and to cross-analyze the presented findings in this paper with findings from future case studies.

\bibliographystyle{plain}
\bibliography{oireferences}

\begin{thebibliography}{10}

\bibitem{alspaugh2013ongoing}
Thomas Alspaugh and Walt Scacchi.
\newblock Ongoing software development without classical requirements.
\newblock In {\em 2013 21st IEEE International Requirements Engineering
  Conference (RE)}, pages 165--174, July 2013.

\bibitem{assink2006inhibitors}
Marnix Assink.
\newblock Inhibitors of disruptive innovation capability: a conceptual model.
\newblock {\em European Journal of Innovation Management}, 9(2):215--233, 2006.

\bibitem{beck2001agile}
Kent Beck, Mike Beedle, Arie van Bennekum, Alistair Cockburn, Ward Cunningham,
  Martin Fowler, James Grenning, Jim Highsmith, Andrew Hunt, Ron Jeffries,
  et~al.
\newblock Manifesto for agile software development.
\newblock \url{http://agilemanifesto.org/principles.html}, 2001.

\bibitem{bird2012examining}
Christian Bird and Nachiappan Nagappan.
\newblock Who? where? what?: Examining distributed development in two large
  open source projects.
\newblock In {\em Proceedings of the 9th IEEE Working Conference on Mining
  Software Repositories}, MSR '12, pages 237--246, Piscataway, NJ, USA, 2012.
  IEEE Press.

\bibitem{Bjarnason2015}
Elizabeth Bjarnason, Michael Unterkalmsteiner, Emelie Engstr{\"o}m, and Markus
  Borg.
\newblock An industrial case study on test cases as requirements.
\newblock In {\em Agile Processes in Software Engineering and Extreme
  Programming: 16th International Conference, XP 2015, Helsinki, Finland, May
  25-29, 2015, Proceedings}, pages 27--39, Cham, 2015. Springer International
  Publishing.

\bibitem{chesbrough2006}
Henry Chesbrough, Wim Vanhaverbeke, and Joel West.
\newblock {\em Open innovation: Researching a new paradigm}.
\newblock Oxford University Press, 2006.

\bibitem{chesbrough2014}
Henry Chesbrough, Wim Vanhaverbeke, and Joel West, editors.
\newblock {\em New Frontiers in Open Innovation}.
\newblock Oxford University Press, 2014.

\bibitem{chesbrough_open_2003}
Henry~William Chesbrough.
\newblock {\em Open innovation: the new imperative for creating and profiting
  from technology}.
\newblock Harvard Business School Press, Boston, Mass., 2003.

\bibitem{chesbrough2007open}
Henry~William Chesbrough and Melissa~M. Appleyard.
\newblock Open innovation and strategy.
\newblock {\em California Management Review}, 50(1):57--76, 2007.

\bibitem{Cruzes2011440}
Daniela~S. Cruzes and Tore Dyb{\aa}.
\newblock Research synthesis in software engineering: A tertiary study.
\newblock {\em Information and Software Technology}, 53(5):440 -- 455, 2011.

\bibitem{Cruzes14}
Daniela~S. Cruzes, Tore Dyb{\aa}, Per Runeson, and Martin H{\"o}st.
\newblock Case studies synthesis: a thematic, cross-case, and narrative
  synthesis worked example.
\newblock {\em Empirical Software Engineering}, 20(6):1634--1665, 2015.

\bibitem{Dahlander2008}
Linus Dahlander and Mats Magnusson.
\newblock How do firms make use of open source communities?
\newblock {\em Long Range Planning}, 41(6):629 -- 649, 2008.

\bibitem{Dahlander2005481}
Linus Dahlander and Mats~G. Magnusson.
\newblock Relationships between open source software companies and communities:
  Observations from nordic firms.
\newblock {\em Research Policy}, 34(4):481 -- 493, 2005.

\bibitem{Dahlander20061243}
Linus Dahlander and Martin Wallin.
\newblock A man on the inside: Unlocking communities as complementary assets.
\newblock {\em Research Policy}, 35(8):1243 -- 1259, 2006.

\bibitem{daniel2011}
Sherae Daniel, Likoebe Maruping, Marcelo Cataldo, and James Herbsleb.
\newblock When cultures clash: {Participation} in open source communities and
  its implications for organizational commitment.
\newblock In {\em Proceedings of the International Conference on Information
  Systems (ICIS)}, Shanghai, China, 2011.

\bibitem{Edison2013}
Henry Edison, Nauman Bin~Ali, and Richard Torkar.
\newblock Towards innovation measurement in the software industry.
\newblock {\em Journal of Systems and Software}, 86(5):1390 -- 1407, 2013.

\bibitem{Enkel2009}
Ellen Enkel, Oliver Gassmann, and Henry Chesbrough.
\newblock Open r{\&}d and open innovation: exploring the phenomenon.
\newblock {\em R{\&}D Management}, 39(4):311--316, 2009.

\bibitem{ernst2012case}
Neil Ernst and George Murphy.
\newblock Case studies in just-in-time requirements analysis.
\newblock In {\em 2012 Second IEEE International Workshop on Empirical
  Requirements Engineering (EmpiRE)}, pages 25--32, Sept 2012.

\bibitem{fricker2010requirements}
Samuel Fricker.
\newblock Requirements value chains: Stakeholder management and requirements
  engineering in software ecosystems.
\newblock In {\em Requirements Engineering: Foundation for Software Quality},
  pages 60--66. Springer, 2010.

\bibitem{gassmann_towards_2004}
Oliver Gassmann and Ellen Enkel.
\newblock Towards a theory of open innovation: three core process archetypes.
\newblock In {\em Proceedings of the R\&D Management Conference}, pages 1--18,
  Lisbon, Portugal, 2004.

\bibitem{gonzalez2013understanding}
Jesus~M Gonzalez-Barahona, Daniel Izquierdo-Cortazar, Stefano Maffulli, and
  Gregorio Robles.
\newblock Understanding how companies interact with free software communities.
\newblock {\em IEEE software}, 30(5):38--45, 2013.

\bibitem{grotnes2009standardization}
Endre Gr{\o}tnes.
\newblock Standardization as open innovation: two cases from the mobile
  industry.
\newblock {\em Information Technology \& People}, 22(4):367--381, 2009.

\bibitem{hattorinature2008}
Lile Hattori and Michele Lanza.
\newblock On the nature of commits.
\newblock In {\em 2008 23rd IEEE/ACM International Conference on Automated
  Software Engineering - Workshops}, pages 63--71, Sept 2008.

\bibitem{henkel_selective_2006}
Joachim Henkel.
\newblock Selective revealing in open innovation processes: The case of
  embedded linux.
\newblock {\em Research Policy}, 35(7):953--969, 2006.

\bibitem{henkelchampions2008}
Joachim Henkel.
\newblock Champions of revealing-the role of open source developers in
  commercial firms.
\newblock {\em Industrial and Corporate Change}, 18(3):435--471, December 2008.

\bibitem{henkel_emergence_2013}
Joachim Henkel, Simone Schöberl, and Oliver Alexy.
\newblock The emergence of openness: How and why firms adopt selective
  revealing in open innovation.
\newblock {\em Research Policy}, 43(5):879 -- 890, 2014.
\newblock Open Innovation: New Insights and Evidence.

\bibitem{Gerrit}
Google~Project Hosting.
\newblock Gerrit code review source code repository.
\newblock \url{https://code.google.com/p/gerrit/wiki/Source?tm=4}.
\newblock Accessed: 2014-06-24.

\bibitem{Husig2011}
Stefan Huesig and Stefan Kohn.
\newblock Open cai 2.0 - computer aided innovation in the era of open
  innovation and web 2.0.
\newblock {\em Computers in Industry}, 62(4):407 -- 13, 2011.

\bibitem{Jansen2012SGO}
Slinger Jansen, Sjaak Brinkkemper, Jurriaan Souer, and Lutzen Luinenburg.
\newblock Shades of gray: Opening up a software producing organization with the
  open software enterprise model.
\newblock {\em Journal of Systems and Software.}, 85(7):1495--1510, July 2012.

\bibitem{jensen2007role}
Chris Jensen and Walt Scacchi.
\newblock Role migration and advancement processes in ossd projects: A
  comparative case study.
\newblock In {\em 29th International Conference on Software Engineering
  (ICSE'07)}, pages 364--374, May 2007.

\bibitem{jensen2010governance}
Chris Jensen and Walt Scacchi.
\newblock Governance in open source software development projects: A
  comparative multi-level analysis.
\newblock In {\em IFIP International Conference on Open Source Systems}, pages
  130--142. Springer, 2010.

\bibitem{knauss2014openness}
Eric Knauss, Daniela Damian, Alessia Knauss, and Arber Borici.
\newblock Openness and requirements: Opportunities and tradeoffs in software
  ecosystems.
\newblock In {\em 2014 IEEE 22nd International Requirements Engineering
  Conference (RE)}, pages 213--222, Aug 2014.

\bibitem{lakhani_principles_2007}
Karim Lakhani and Jill~A. Panetta.
\newblock The principles of distributed innovation.
\newblock {SSRN} Scholarly Paper {ID} 1021034, Social Science Research Network,
  Rochester, {NY}, October 2007.

\bibitem{Lakhani2003923}
Karim~R Lakhani and Eric von Hippel.
\newblock How open source software works: “free” user-to-user assistance.
\newblock {\em Research Policy}, 32(6):923 -- 943, 2003.

\bibitem{Lee2003}
Gwendolyn~K. Lee and Robert~E. Cole.
\newblock From a firm-based to a community-based model of knowledge creation:
  The case of the linux kernel development.
\newblock {\em Organization Science}, 14(6):633--649, 2003.

\bibitem{Lee2009426}
Sang-Yong~Tom Lee, Hee-Woong Kim, and Sumeet Gupta.
\newblock Measuring open source software success.
\newblock {\em Omega}, 37(2):426 -- 438, 2009.

\bibitem{lerner2002some}
Josh Lerner and Jean Tirole.
\newblock Some simple economics of open source.
\newblock {\em The Journal of Industrial Economics}, 50(2):197--234, 2002.

\bibitem{lieberman1998first}
Marvin~B Lieberman and David~Bruce Montgomery.
\newblock {\em First-mover (dis) advantages: Retrospective and link with the
  resource-based view}.
\newblock Graduate School of Business, Stanford University, 1998.

\bibitem{linaaker2014}
Johan Lin{\aa}ker, Maria Krantz, and Martin H{\"o}st.
\newblock On infrastructure for facilitation of inner source in small
  development teams.
\newblock In {\em Product-Focused Software Process Improvement}, pages
  149--163. Springer, 2014.

\bibitem{LinakerSurvey15}
Johan Lin{\aa}ker, Husan Munir, Per Runeson, Bj{\"o}rn Regnell, and Claes
  Schrewelius.
\newblock {\em A Survey on the Perception of Innovation in a Large
  Product-Focused Software Organization}, pages 66--80.
\newblock Springer International Publishing, Cham, 2015.

\bibitem{linaaker2016firms}
Johan Lin{\aa}ker, Patrick Rempel, Bj{\"o}rn Regnell, and Patrick M{\"a}der.
\newblock How firms adapt and interact in open source ecosystems: Analyzing
  stakeholder influence and collaboration patterns.
\newblock In {\em Requirements Engineering: Foundation for Software Quality},
  pages 63--81. Springer, 2016.

\bibitem{lindencommodification2009}
Frank van~der Linden, Bj\"orn Lundell, and Pentti Marttiin.
\newblock Commodification of industrial software: A case for open source.
\newblock {\em {IEEE} Software}, 26(4):77--83, 2009.

\bibitem{lindman2008applying}
Juho Lindman, Matti Rossi, and Pentti Marttiin.
\newblock Applying open source development practices inside a company.
\newblock In {\em Open Source Development, Communities and Quality}, pages
  381--387. Springer, 2008.

\bibitem{cvsanaly}
MetricsGrimoire.
\newblock {CVSAnalY}.
\newblock \url{http://metricsgrimoire.github.io/CVSAnalY/}.
\newblock Accessed: 2014-07-17.

\bibitem{mockuswhy2002}
Audris Mockus and James~D. Herbsleb.
\newblock Why not improve coordination in distributed software development by
  stealing good ideas from open source.
\newblock In {\em Meeting {Challenges} and {Surviving} {Success}: {The} 2nd
  {Workshop} on {Open} {Source} {Software} {Engineering}}, pages 19--25, 2002.

\bibitem{mollercritical2015}
Charlotte M\"{o}ller and Madeleine Wahlqvist.
\newblock Critical {Success} {Factors} for {Innovative} {Performance} of
  {Individuals}.
\newblock {\em Management}, 39(5):1155--1161, 2012.

\bibitem{Morgan2012}
Lorraine Morgan, Joseph Feller, and Patrick Finnegan.
\newblock Exploring inner source as a form of intra-organisational open
  innovation.
\newblock In {\em 19th European Conference on Information Systems (ECIS)},
  Helsinki, Finland, 2011.

\bibitem{Morgan2010}
Lorraine Morgan and Patrick Finnegan.
\newblock Open innovation in secondary software firms: An exploration of
  managers' perceptions of open source software.
\newblock {\em SIGMIS Database}, 41(1):76--95, February 2010.

\bibitem{mowery_plus_2009}
David~C. Mowery.
\newblock Plus ca change: Industrial {R\&D} in the third industrial revolution.
\newblock {\em Industrial and Corporate Change}, 18(1):1--50, 2009.

\bibitem{RedHatIBM2009}
Neeshal Munga, Thomas Fogwill, and Quentin Williams.
\newblock The adoption of open source software in business models: A red hat
  and ibm case study.
\newblock In {\em Proceedings of the 2009 Annual Research Conference of the
  South African Institute of Computer Scientists and Information
  Technologists}, SAICSIT '09, pages 112--121, New York, NY, USA, 2009. ACM.

\bibitem{Munir2014}
Hussan Munir, Misagh Moayyed, and Kai Petersen.
\newblock Considering rigor and relevance when evaluating test driven
  development: A systematic review.
\newblock {\em Information and Software Technology}, 56(4):375 -- 394, 2014.

\bibitem{MunirMapping15}
Hussan Munir, Krzysztof Wnuk, and Per Runeson.
\newblock Open innovation in software engineering: a systematic mapping study.
\newblock {\em Empirical Software Engineering}, 21(2):684--723, 2016.

\bibitem{osloManual2005}
OECD.
\newblock {\em Oslo Manual -- Guidelines for collecting and interpreting
  innovation data}.
\newblock OECD and Eurostat, 3rd edition, 2005.

\bibitem{Jenkins}
Ohloh.net.
\newblock The jenkins gerrit trigger plugin open source project.
\newblock \url{https://www.ohloh.net/p/gerrit-trigger-plugin}.
\newblock Accessed: 2014-07-08.

\bibitem{panjer2008cooperation}
Lucas~D Panjer, Daniela Damian, and Margaret-Anne Storey.
\newblock Cooperation and coordination concerns in a distributed software
  development project.
\newblock In {\em Proceedings of the 2008 international workshop on Cooperative
  and human aspects of software engineering}, pages 77--80. ACM, 2008.

\bibitem{Pohl2005}
Klaus Pohl, G\"{u}nter B\"{o}ckle, and Frank J. van~der Linden.
\newblock {\em Software Product Line Engineering: Foundations, Principles and
  Techniques}.
\newblock Springer-Verlag New York, Inc., Secaucus, NJ, USA, 2005.

\bibitem{Rolandsson2011576}
Bertil Rolandsson, Magnus Bergquist, and Jan Ljungberg.
\newblock Open source in the firm: Opening up professional practices of
  software development.
\newblock {\em Research Policy}, 40(4):576 -- 587, 2011.

\bibitem{Runesoncasestudy2012}
Per Runeson, Martin H{\"o}st, Austen Rainer, and Bj{\"o}rn Regnell.
\newblock {\em Case Study Research in Software Engineering - Guidelines and
  Examples}.
\newblock Wiley, 2012.

\bibitem{scacchi2002}
Walt Scacchi.
\newblock Understanding the requirements for developing open source software
  systems.
\newblock In {\em Software, IEE Proceedings-}, volume 149, pages 24--39. IET,
  2002.

\bibitem{scacchi2010collaboration}
Walt Scacchi.
\newblock Collaboration practices and affordances in free/open source software
  development.
\newblock In {\em Collaborative software engineering}, pages 307--327.
  Springer, 2010.

\bibitem{staahl2014modeling}
Daniel St{\aa}hl and Jan Bosch.
\newblock Modeling continuous integration practice differences in industry
  software development.
\newblock {\em Journal of Systems and Software}, 87:48--59, 2014.

\bibitem{Stam2009}
Wouter Stam.
\newblock When does community participation enhance the performance of open
  source software companies?
\newblock {\em Research Policy}, 38(8):1288 -- 1299, 2009.

\bibitem{stol2014key}
Klaas-Jan Stol, Paris Avgeriou, Muhammad~Ali Babar, Yan Lucas, and Brian
  Fitzgerald.
\newblock Key factors for adopting inner source.
\newblock {\em ACM Transactions on Software Engineering and Methodology
  (TOSEM)}, 23(2):18, 2014.

\bibitem{stuermer_extending_2009}
Matthias Stuermer, Sebastian Spaeth, and Georg Von~Krogh.
\newblock Extending private-collective innovation: a case study.
\newblock {\em R\&D Management}, 39(2):170--191, 2009.

\bibitem{ven2008challenges}
Kris Ven and Herwig Mannaert.
\newblock Challenges and strategies in the use of open source software by
  independent software vendors.
\newblock {\em Information and Software Technology}, 50(9):991--1002, 2008.

\bibitem{wesselius2008}
Jacco Wesselius.
\newblock The bazaar inside the cathedral: Business models for internal
  markets.
\newblock {\em IEEE Software}, 25(3):60--66, May 2008.

\bibitem{West20031259}
Joel West.
\newblock How open is open enough?: Melding proprietary and open source
  platform strategies.
\newblock {\em Research Policy}, 32(7):1259 -- 1285, 2003.

\bibitem{West2013}
Joel West and Marcel Bogers.
\newblock Leveraging external sources of innovation: A review of research on
  open innovation.
\newblock {\em Journal of Product Innovation Management}, 31(4):814--831, 2014.

\bibitem{west2006challenges}
Joel West and Scott Gallagher.
\newblock Challenges of open innovation: the paradox of firm investment in
  open-source software.
\newblock {\em R\&D Management}, 36(3):319--331, 2006.

\bibitem{west_creating_2008}
Joel West and David Wood.
\newblock Creating and evolving an open innovation ecosystem: Lessons from
  symbian ltd.
\newblock {\em Available at {SSRN} 1532926}, 2008.

\bibitem{west2013evolving}
Joel West and David Wood.
\newblock Evolving an open ecosystem: The rise and fall of the symbian
  platform.
\newblock {\em Advances in Strategic Management}, 30:27--67, 2013.

\bibitem{Wnuk2012}
Krzysztof Wnuk, Dietmar Pfahl, David Callele, and Even-Andr{\'e} Karlsson.
\newblock How can open source software development help requirements management
  gain the potential of open innovation: An exploratory study.
\newblock In {\em Proceedings of the ACM-IEEE International Symposium on
  Empirical Software Engineering and Measurement}, ESEM '12, pages 271--280,
  New York, NY, USA, 2012. ACM.

\end{thebibliography}

\newpage
\appendix
\label{sec:Appendix}
\section{APPENDIX: Supplementary interview questionnaire}
  \begin{longtable}{p{11cm}}
    \toprule
    \multicolumn{1}{l}{\textbf{Demographics} } \\
    \midrule 
    \begin{itemize}
    \item Where do you work? 
    \item What is your job title?
    \item Which department do you work for in the organization?
    \item How many years of experience do you have?
    \item Could you, in short, describe your daily work and responsibilities?
    \end{itemize}

     \\
      \midrule
      \textbf{General involvement}  \\
    \midrule
   
     \begin{itemize}
     \item Are you, or have been, in any way actively involved in any open source community in your daily work? (Gerrit, Jenkins, any other?)
     \item Could you describe your involvement?
     \item What is/was the reasons for your involvement in these open source communities? (Volunteered or tasked by management?) 
     \item How much time are you allowed to spend on community interaction?
     \item How is your involvement with these community in your spare time, outside of your daily work?
     \item What development process/methodology do you use and how does it interact with the community? (process of working)
     \end{itemize}
  \\
      \midrule
     \textbf{Requirements}  \\
    \midrule
    \begin{itemize}
    \item What are the sources (internal and external) behind the requirements/features? (by tool developers, tool users, pm's, others…)
    \item How do you manage and implement the requirements/features?
    \item How are the requirements approved and prioritized? (By developers alone, pm's, community…)
    \item How is your involvement perceived from the community? Positive or negative? How come? (competitors)
    \item Are there any internal (organizational) obstacles in contributing to the community? (Time, IP, management….)
    \item Are there any external obstacles related to the involvement in the community related to the addition of new requirements/features?
    \item How did you overcome these?
    \end{itemize}
   \textbf{} \\
          \midrule
      \textbf{Testing}  \\
     \midrule
     \begin{itemize}
     \item How does your internal process of reporting bugs differ from the community's? (tools for reporting bugs in community)
     \item How do you manage traceability between tests and requirements?
     \item Who is responsible for fixing those bugs? (Process behind, consequence on quality and resolution time)
     \item How does your internal process for correcting bugs or issues, differ from the community's?
	 \item Are there any obstacles related to the involvement in the community related to the testing process? How did you overcome these? (Communication, synchronized level of quality/tests between contributors)
    
     \end{itemize}
     \\
          \midrule
    
      \textbf{Business/strategy}  \\
    \midrule
        \begin{itemize}
    \item What motivates your organization to contribute to open source project(s)? (Beyond lower cost, improved quality?)
    \item What is the strategy behind these commits? 
    \item Did you consider alternate strategies such as buying proprietary tools (COTS) or hiring people/outsourcing for the development these tools? Why?
    \item How are these strategies supported by your internal procedures (IP department)? 
    \item Is it a local strategy or global strategy? Who are the sponsors? 
    \item How has the commits effected the relation with other (corporate) stakeholders in the communities? (Free-riding, governance structure, constraints, Sony Authority, collaboration, balance between community and Sony's needs, community buildup) 
    \item How has the commits effected the relation with other competitors? (Free-riding, governance structure, collaboration) 
    \end{itemize}
    \textbf{}  \\
     \midrule
      \textbf{Perception on innovation and outcome}  \\
        \begin{itemize}
    \item How has the usage/development of these tools effected the Sony Mobile's product development? (Developers, testers) 
    \item How has the usage of these tools effected the products? 
    \item Is innovativeness of a requirement/issue/bug considered, and if so, what effect does it have on the requirements and contribution process? 
    \item How has the involvement in the communities implicated on innovation in your: 1) Processes? 2) Products 3) Organization 4) Business strategies
    \item How do you measure the impact from the development/usage of these tools on Sony Mobile's product development? Metrics etc. 
    \item Is the knowledge gained from the OSS tool development transferred and exploited outside of the tools development? (Absorptive capacity – Firm level, individual level)
    \end{itemize}
  \\
          \midrule
      \textbf{Ending remarks}  \\
    \bottomrule
 \label{tab:interviewquestions}
\end{longtable}

\end{document}